\documentclass[11pt]{article}
\usepackage[utf8]{inputenc}
\usepackage{titling}
\usepackage{authblk}
\usepackage{amsfonts}
\usepackage{amsmath}
\usepackage{amssymb}
\usepackage{epsfig}
\usepackage{graphicx,color}
\usepackage{bm}
\usepackage{cite}
\usepackage{setspace}
\begin{document}
\title{\textbf{The $\bm{\Lambda_b\,\to} \bm{\Lambda\,(\to p \pi^-)\mu^+\mu^-}$ decay in the $\text{RS}_{\bm{c}}$ model}}
\author[1]{Aqsa Nasrullah\thanks{aqsanasrullah54@gmail.com}}
\author[2,3]{Faisal Munir Bhutta\thanks{faisalmunir@ihep.ac.cn}}
\author[1]{M. Jamil Aslam\thanks{jamil@qau.edu.pk}}
\affil[1]{Physics Department, Quaid-i-Azam University, Islamabad, Pakistan}
\affil[2]{Institute of High Energy Physics, Chinese Academy of Sciences, Beijing 100049, China}
\affil[3]{University of Chinese Academy of Sciences, Beijing 100049, China}
\renewcommand\Authands{ and }
\predate{}\postdate{}\date{}
\maketitle
\singlespacing
\begin{abstract}
We study the four body decay $\Lambda_b \rightarrow \Lambda ( \rightarrow p \pi^-) \mu^{+} \mu^{-}$ in the Randall-Sundrum model
with custodial protection $(\text{RS}_c)$. By considering the constraints coming from the direct searches of the lightest
Kaluza-Klein (KK) excitation of the gluon, electroweak precision tests, the measurements of the
Higgs signal strengths at the LHC and from $\Delta F=2$ flavor observables, we perform a scan of the parameter space of
the $\text{RS}_c$ model and obtain the maximum allowed deviations
of the Wilson coefficients $\Delta C^{(\prime)}_{7,\; 9,\; 10}$ for different values of the lightest KK gluon mass $M_{g^{(1)}}$.
Later, their implications on the observables such as differential branching fraction, longitudinal polarization of the
daughter baryon $\Lambda$, forward-backward asymmetry with respect to leptonic, hadronic and combined lepton-hadron angles are
discussed where we present the analysis of these observables in different bins of di-muon invariant mass squared $s\;(= q^2)$.
It is observed that with the current constraints the Wilson coefficients in $\text{RS}_c$ model show slight deviations
from their Standard Model values and hence can not accommodate the discrepancies between the Standard Model
calculations of various observables and the LHCb measurements in $\Lambda_b$ decays.
\end{abstract}
\onehalfspacing
\section{Introduction\label{Intro}}
Although the Large Hadron Collider (LHC) has so far not observed any new particles directly, that are predicted by many beyond Standard Model (SM) scenarios, it has certainly provided some intriguing discrepancies from the SM expectations in semi-leptonic rare $B$-meson decays. In this context,
a persistent pattern of deviations in tension with the SM predictions has been emerging from observables in a number of
$b\to sl^+l^-$ processes. In particular, LHCb measurements \cite{Aaij:2014ora,Aaij:2017vbb} of the observables
$R_K$ and $R_{K^{\ast}}$ representing the ratios of branching fractions $B^+\to K^+\mu^+\mu^-$ to $B^+\to K^+e^+e^-$ and
$B^0\to K^{\ast0}\mu^+\mu^-$ to $B^0\to K^{\ast0}e^+e^-$, respectively, show deviations from the SM predictions $\sim 1$ and
together they indicate the lepton flavor universality violation with the significance at the $4\sigma$ level
\cite{Altmannshofer:2017yso,DAmico:2017mtc,Geng:2017svp,Hiller:2017bzc}. Further, the LHCb results for the branching
fractions of the $B\to K^{(\ast)}\mu^+\mu^-$ and $B_s\to \phi\mu^+\mu^-$ decays \cite{Aaij:2014pli,Aaij:2013aln,Aaij:2015esa},
suggest the smaller values compared to their SM estimates. Moreover, mismatch between the LHCb findings and the SM predictions
in the angular analysis of the $B^0\to K^{\ast0}\mu^+\mu^-$ decay \cite{Aaij:2013qta,Aaij:2015oid}, with the confirmation by the
Belle collaboration later on \cite{Wehle:2016yoi}, has become a longstanding issue. In this context, recent phenomenological analyses
have explored the underlying new physics (NP) possibilities behind these anomalies \cite{Altmannshofer:2017yso,DAmico:2017mtc,Geng:2017svp,Hiller:2017bzc,
Ciuchini:2017mik,Hurth:2017hxg,Chiang:2017hlj,Capdevila:2017bsm,Kohda:2018xbc,Falkowski:2018dsl}. However, to establish the claim that the deviations in the angular asymmetries in $B \to K^{*}(\to K\pi) \mu^{+}\mu^{-}$ decays are indications of NP, an improvement is needed both on the theoretical and the experimental sides. On theoretical front we have to get better control on the hadronic uncertainties arising mainly due to form factors (FF) and on the experimental end, some more data with improved statistics is needed which is expected from the Belle II and LHCb.
Another possibility that exist on the theoretical side is to analyze more processes which are mediated by the same quark level transition $b \to s \mu^+\mu^-$.

Among them, the rare baryonic decay $\Lambda_b \to \Lambda \mu^+\mu^-$ is particularly important as it can provide complementary information
and additionally offers a unique opportunity to understand the helicity structure of the effective weak Hamiltonian for
$b \to s$ transition \cite{Mannel:1997xy,Chen:2001zc}. The branching ratio for this decay was first measured by CDF collaboration \cite{Aaltonen:2011qs}. Recently,  the LHCb has reported its measurements for branching ratio and three angular
observables \cite{Aaij:2015xza} in the
$\Lambda_b \to \Lambda(\to p\pi^-) \mu^+\mu^-$ decay. Theoretically
challenging aspect in the study of the $\Lambda_b \to \Lambda \mu^+\mu^-$ decay is the evaluation of the hadronic $\Lambda_b\to \Lambda$ transition from
factors. In this context, recent progress is made by performing the high precision lattice QCD calculations \cite{Detmold:2016pkz}.
Moreover, these FF have been estimated using various models or approximations such as quark models \cite{Cheng:1995fe,Mott:2011cx},
perturbative QCD \cite{He:2006ud}, SCET \cite{Feldmann:2011xf} and QCD light cone sum-rules (LCSR) \cite{Wang:2008sm,Wang:2009hra,Wang:2015ndk}.
Furthermore, extensive studies of the semi-leptonic decays of $\Lambda_b$ baryon $(\Lambda_b \to \Lambda \ell^+\ell^-)$, both within the SM and in many different NP scenarios, have been performed \cite{Aliev:2002hj,Aliev:2002tr,Aliev:2004yf,Turan:2005cw,Giri:2005mt,Aliev:2006xd,Bashiry:2007pd,Aslam:2008hp,Wang:2008ni,Aliev:2010uy,Azizi:2010qk,Azizi:2011ri
,Azizi:2012vy,Gutsche:2013pp,Azizi:2013eta,Paracha:2014hca,Boer:2014kda,Meinel:2016grj,Wang:2016dne,Azizi:2016dcj,Hu:2017qxj,Faustov:2017wbh,Alnahdi:2017ogx,
Roy:2017dum,Das:2018sms,Nasrullah:2018puc}. Recently, the angular distributions for polarized $\Lambda_b$ are presented in \cite{Blake:2017une}.

In the present work, we study the four body $\Lambda_b \to \Lambda(\to p\pi^-) \mu^+\mu^-$ decay in the framework of the Randall-Sundrum (RS) model
with custodial protection. The RS model features five-dimensional (5D) space-time with a non-trivial warped metric \cite{Randall:1999ee}.
After performing the KK decomposition and integrating over the fifth dimension the effective 4D theory is obtained which involves new particles appearing
as the KK resonances, either of the SM particles or the ones which do not possess SM counterparts. Assuming that the weak effective Hamiltonian of the
$\Lambda_b \to \Lambda(\to p\pi^-) \mu^+\mu^-$ decay emerges from the well-defined theory of the $\text{RS}_c$ model, the Wilson coefficients of the
effective Hamiltonian get modified with respect to the SM values due to additional contributions from the heavy KK excitations and are correlated in a unique way.
Expecting distinct phenomenological consequences from such a correlation on the angular observables of the $\Lambda_b \to \Lambda(\to p\pi^-) \mu^+\mu^-$ decay,
we study whether the current experimental data on this decay can be explained in the $\text{RS}_c$ model.

Although $B$-meson decays have been investigated extensively in different variants of the RS model \cite{Burdman:2003nt,Agashe:2004cp,Moreau:2006np,Casagrande:2008hr,Blanke:2008zb,Blanke:2008yr,Albrecht:2009xr,Bauer:2009cf,Blanke:2012tv,Biancofiore:2014wpa,
Biancofiore:2014uba,Lu:2016pfs,DAmbrosio:2017wis,Blanke:2018sro}, not many studies are devoted to the $\Lambda_b$ decays
in the RS model \cite{Azizi:2015hoa}. Additionally, our present study includes new considerations and results which were not available in the
previous studies of the $\Lambda_b$ decays entertaining the RS model. Firstly, we will consider the current
constraints on the parameter space of the $\text{RS}_c$ model coming from the direct searches of the lightest KK gluon, electroweak
precision tests and from the measurements of the Higgs signal strengths at the LHC, which yield much stricter
constraints on the mass scale of the lowest KK gluon $M_{g^{(1)}}$, which in turn prevent sizeable deviations of the Wilson coefficients
from the SM predictions. Secondly, we will not adopt the simplification of treating the elements of the
5D Yukawa coupling matrices to be real numbers as considered in \cite{Azizi:2015hoa,Biancofiore:2014wpa}, rather we will take these entries to
be complex numbers as considered in \cite{Lu:2016pfs,Blanke:2008zb} leading to the complex Wilson coefficients instead of real ones. Last but not
the least, we will use the helicity parametrization of the $\Lambda_b \to \Lambda$ hadronic matrix elements and for the involved FF, we
will use the most recent lattice QCD calculations, both in the low and high $q^2$ regions, which yield much smaller uncertainties in most of the kinematic range \cite{Detmold:2016pkz}.

The rest of the paper is organized as follows. In Sec. \ref{RScmodel}, we describe the essential features of the $\text{RS}_c$ model especially relevant
for the study of the considered decay. In Sec. \ref{setup}, we present the theoretical formalism including the effective weak Hamiltonian, analytical
expressions of the Wilson coefficients in the $\text{RS}_c$ model and the angular observables of interest in the four-body
$\Lambda_b \to \Lambda(\to p\pi^-) \mu^+\mu^-$ decay. After discussing the current constraints and subsequently scanning the
parameter space of the $\text{RS}_c$ model in Sec. \ref{Rscbounds}, we give our numerical results and their discussion in Sec. \ref{num}.
Finally, in Sec. \ref{conc}, we conclude our findings.

\section{RS Model with Custodial Symmetry\label{RScmodel}}
In this section we will describe some of the salient features of the RS model \cite{Randall:1999ee}.
The RS model, also known as warped extra dimension, offers a
geometrical solution of the gauge hierarchy problem along with naturally
explaining the observed hierarchies in the SM fermion masses and mixing angles.
The model is described in a five-dimensional space-time, where the fifth
dimension is compactified on an orbifold and the non-factorizable RS metric is
given by
\begin{eqnarray}\label{01}
ds^2=e^{-2ky}\eta_{\mu\nu}dx^{\mu}dx^{\nu}-dy^2,
\end{eqnarray}
where $k\sim \mathcal{O}(M_{\text{Pl}})\simeq10^{19}$ GeV is the curvature scale, $\eta_{\mu\nu}=\text{diag}(+1,-1,-1,-1)$ is the 4D Minkowski
metric and $y$ is the extra-dimensional (fifth) coordinate which varies
in the finite interval $0\le y\le L$; the endpoints of the interval $y=0$ and
$y=L$ represent the boundaries of the extra dimension and are known as \textit{ultraviolet}
(UV) and \textit{Infrared} (IR) brane, respectively. The region in between the UV and IR brane
is denoted as the bulk of the warped extra dimension. In order to solve the gauge hierarchy
problem, we take $kL=36$ and define
\begin{eqnarray}\label{02}
M_{\text {KK}}\equiv ke^{-kL}\sim\mathcal{O}(\text{TeV}),
\end{eqnarray}
as the only free parameter coming from space-time geometry representing the effective NP scale.

In the present study, we consider a specific setup of the RS model in which the SM gauge group is
enlarged to the bulk gauge group
\begin{eqnarray}\label{gaugegroup}
SU(3)_c\times SU(2)_L\times SU(2)_R\times U(1)_X\times P_{LR},
\end{eqnarray}
which is known as the RS model with custodial protection $(\text {RS}_c)$
\cite{Agashe:2006at,Carena:2006bn,Contino:2006qr,Cacciapaglia:2006gp,Albrecht:2009xr}. $P_{LR}$ is
the discrete symmetry, interchanging the two $SU(2)_{L,R}$ groups, which is responsible
for the protection of the $Zb_L\bar{b}_L$ vertex. Moreover, for this particular scenario
it has been shown that all existing $\Delta F=2$ and electroweak (EW) precision
constraints can be satisfied, without requiring too much fine-tuning, for
the masses of the lightest KK excitations of the order of a few
TeV \cite{Blanke:2008zb}, in the reach of the LHC. However, after the ATLAS and the CMS
measurements of the Higgs signal strengths, the bounds on the masses of the lightest
KK modes arising from Higgs physics have grown much stronger than those stemming from
EW precision measurements \cite{Malm:2014gha}. In view of this, we have performed a scan for the allowed parameter
space of the model by considering all existing constraints, which will be discussed later on.

In the chosen setup, all the SM fields are allowed to propagate in the 5D bulk, except the
Higgs field, which is localized near or on the IR brane. In the present study we consider
the case in which Higgs boson is completely localized on the IR brane at $y=L$. The $\text {RS}_c$ model features
two symmetry breakings. First, the enlarged gauge group of the model is broken down to the SM gauge group after
imposing suitable boundary conditions (BCs) on the UV brane. Later on the spontaneous symmetry breaking occurs through
Higgs mechanism on the IR brane. As a natural consequence in all the extra dimensional models, we have an infinite tower of KK excitations in this model.
For this, each 5D field $F(x^{\mu},y)$ is KK decomposed to generic form
\begin{eqnarray}\label{03}
F(x^{\mu},y)=\frac{1}{\sqrt{L}} \sum_{n=0}^{\infty} F^{(n)}(x^{\mu})f^{(n)}(y),
\end{eqnarray}
where $F^{(n)}(x^{\mu})$ represent the effective four-dimensional fields and $f^{(n)}(y)$ are called as the
five-dimensional profiles or the shape functions. $n=0$ case, called as zero mode in the KK mode expansion of a given field, corresponds to the
SM particle. Appropriate choices for BCs help to distinguish between fields with and without a zero mode. Fields with
the Neumann BCs on both branes, denoted as $(++)$, have a zero mode that can be identified with a SM particle while fields
with the Dirichlet BC on the UV brane and Neumann BC on the IR brane, denoted as $(-+)$, do not have the SM partners. Profiles for
different fields are obtained by solving the corresponding 5D bulk equations of motion (EOM). In a perturbative approach as described
in \cite{Albrecht:2009xr}, EOMs can be solved before the electroweak symmetry breaking (EWSB) and after the Higgs field develops
a vacuum expectation value (VEV), the ratio $\upsilon/M_{g^{(1)}}$ of the Higgs VEV $\upsilon$ and the mass of the lowest KK excitation mode of gauge bosons $M_{g^{(1)}}$ can be taken as perturbation.\footnote{Here we mention that we have employed a different notation for the mass of the
first KK gauge bosons than in \cite{Albrecht:2009xr} such that our $M_{\text {KK}}$ corresponds to their $f$.} Starting with the action of 5D theory, we integrate over the fifth dimension $y$ to obtain the 4D effective field theory, and the Feynman rules of the model are obtained by neglecting terms of $\mathcal{O}(\upsilon^{2}/M_{g^{(1)}}^2)$ or higher. On similar grounds,
the mixing occurring between the SM fermions and the higher
KK fermion modes can be neglected as it leads to $\mathcal{O}(\upsilon^{2}/M_{g^{(1)}}^2)$ modifications of the relevant couplings.


Next, we discuss the particle content of the gauge sector of the $\text{RS}_c$ model and the mixing between SM gauge bosons and the first higher KK modes
after the EWSB. For gauge bosons, following the analyses performed in Refs. \cite{Blanke:2008zb,Biancofiore:2014wpa}, we have neglected the $n>1$ KK modes as it is
observed that the model becomes non-perturbative already for scales corresponding to the first few KK modes. Corresponding
to the enlarged gauge group of the model we have a large number of gauge bosons. For $SU(3)_c$, we have $G_{\mu}^A\; (A=1,...,8)$ corresponding to the SM gluons with 5D coupling $g_s$. The gauge bosons corresponding to $SU(2)_L$ and $SU(2)_R$ are denoted as $W_{L\mu}^a$, and $W_{R\mu}^a\; (a=1,2,3)$
respectively, with 5D gauge coupling $g$. Where the equality of the $SU(2)_L$ and $SU(2)_R$ couplings is imposed by $P_{LR}$ symmetry. The gauge field
corresponding to $U(1)_X$ is denoted as $X_{\mu}$ with 5D coupling $g_X$. All 5D gauge couplings are dimensionful and the relation between 5D and
its 4D counterpart is given by $g_s^{4D}=g_s/\sqrt L$, with similar expressions also existing for $g^{4D}$ and $g_X^{4D}$. Charged gauge bosons
are defined as
\begin{eqnarray}\label{04}
W_{L(R)\mu}^{\pm}=\frac{W_{L(R)\mu}^{1}\mp i W_{L(R)\mu}^{2}}{\sqrt 2}.
\end{eqnarray}
Mixing between the bosons $W_{R\mu}^3$ and $X_{\mu}$ results in fields $Z_{X\mu}$ and $B_{\mu}$,
\begin{eqnarray}\label{05}
Z_{X\mu}&=&\cos \phi\; W_{R\mu}^3-\sin \phi\; X_{\mu},\notag\\
B_{\mu}&=&\sin \phi\; W_{R\mu}^3+\cos \phi\; X_{\mu},
\end{eqnarray}
where
\begin{eqnarray}\label{06}
\cos \phi=\frac{g}{\sqrt{g^2+g_X^2}},\qquad \sin \phi=\frac{g_X}{\sqrt{g^2+g_X^2}}.
\end{eqnarray}
Further, mixing between $W_{L\mu}^3$ and $B_{\mu}$ yields the fields $Z_{\mu}$ and $A_{\mu}$
in analogy to the SM,
\begin{eqnarray}\label{07}
Z_{\mu}&=&\cos \psi\; W_{L\mu}^3-\sin \psi\; B_{\mu},\notag\\
A_{\mu}&=&\sin \psi\; W_{L\mu}^3+\cos \psi\; B_{\mu},
\end{eqnarray}
with
\begin{eqnarray}\label{08}
\cos \psi=\frac{1}{\sqrt{1+\sin^2\phi}},\qquad \sin \psi=\frac{\sin\phi}{\sqrt{1+\sin^2\phi}}.
\end{eqnarray}
Along with eight gluons $G_{\mu}^A(++)$, after the mixing pattern, we have four charged bosons which are specified as
$W_L^{\pm}(++)$ and $W_R^{\pm}(-+)$ while three neutral gauge bosons are given as $A(++)$, $Z(++)$ and $Z_X(-+)$.
Moreover, we mention the following remarks about the masses and profiles of various gauge boson fields that are obtained after solving
the corresponding EOMs. Before EWSB, gauge bosons with $(++)$ BCs have massless zero modes, which correspond to the SM gauge fields, with
flat profiles along the extra dimension. On the other hand gauge bosons with $(-+)$ BCs do not have a zero mode and the lightest
mode in the KK tower starts at $n=1$. The profiles of the first KK mode of gauge bosons having a zero mode are denoted by $g(y)$ and the mass of such modes is
denoted as $M_{++}$ while the first mode profiles of the gauge bosons without a zero mode are given by $\tilde g(y)$ and the
mass of such modes is denoted as $M_{-+}$ before EWSB. There expressions are given by \cite{Gherghetta:2000qt},
\begin{eqnarray}\label{09}
g(y)=\frac{e^{ky}}{N_1}\left[J_1\left(\frac{M_{g^{(1)}}}{k}e^{ky}\right)+b_1(M_{g^{(1)}})Y_1\left(\frac{M_{g^{(1)}}}{k}e^{ky}\right)\right],
\end{eqnarray}
\begin{eqnarray}\label{10}
{\tilde g}(y)=\frac{e^{ky}}{N_1}\left[J_1\left(\frac{{\tilde M}_{g^{(1)}}}{k}e^{ky}\right)+{\tilde b}_1({\tilde M}_{g^{(1)}})Y_1\left(\frac{{\tilde M}_{g^{(1)}}}{k}e^{ky}\right)\right],
\end{eqnarray}
where $J_1$ and $Y_1$ are the Bessel functions of first and second kinds, respectively. The coefficients $b_1(M_{g^{(1)}})\;, {\tilde b}_1({\tilde M}_{g^{(1)}})$ and $N_1$ are
\begin{eqnarray}\label{11}
b_1(M_{g^{(1)}})=-\frac{J_1\left({M_{g^{(1)}}}/{k}\right)+{M_{g^{(1)}}}/{k}J_1^{\prime}\left({M_{g^{(1)}}}/{k}\right)}
{Y_1\left({M_{g^{(1)}}}/{k}\right)+{M_{g^{(1)}}}/{k}Y_1^{\prime}\left({M_{g^{(1)}}}/{k}\right)},
\end{eqnarray}
\begin{eqnarray}\label{12}
{\tilde b}_1({\tilde M}_{g^{(1)}})=-\frac{J_1\left({{\tilde M}_{g^{(1)}}}/{k}\right)}
{Y_1\left({{\tilde M}_{g^{(1)}}}/{k}\right)},
\end{eqnarray}
\begin{eqnarray}\label{13}
N_1=\frac{e^{kL/2}}{\sqrt{\pi LM_{g^{(1)}}}}.
\end{eqnarray}
The masses of the lowest KK gauge excitations are numerically given to be $M_{g^{(1)}}\simeq2.45M_{\text {KK}}\equiv M_{++}$ and
${\tilde M}_{g^{(1)}}\simeq2.40M_{\text {KK}}\equiv M_{-+}$. Notice that the presented KK masses for the gauge bosons are
universal for all gauge bosons with the same BCs. After EWSB, the zero mode gauge bosons
with $(++)$ BCs, other than gluons and photon, acquire masses while the massive KK gauge excitations of all the gauge bosons, except KK gluons and KK photons
receive mass corrections. Due to the unbroken gauge invariance of $SU(3)$ and $U(1)_{Q}$, gluons and photon do not obtain masses such that their zero modes
remain massless while their higher KK excitations that are massive do not get a mass correction as a result of EWSB and hence remain mass eigenstates. Furthermore, we have mixing among zero modes and the higher KK modes. Considering only the first KK modes, the charged and neutral mass eigenstates are related to their corresponding gauge KK eigenstates via
\begin{eqnarray}\label{14}
\left(\begin{array}{c}
W^{\pm} \\
W_H^{\pm} \\
W^{\prime\pm}
\end{array}\right)=\mathcal{G}_W
\left(\begin{array}{c}
W_L^{\pm(0)} \\
W_L^{\pm(1)} \\
W_R^{\pm(1)}
\end{array}\right),
\qquad
\left(\begin{array}{c}
Z \\
Z_H \\
Z^{\prime}
\end{array}\right)=\mathcal{G}_Z
\left(\begin{array}{c}
Z^{(0)} \\
Z^{(1)} \\
Z_X^{(1)}
\end{array}\right).
\end{eqnarray}
The expressions of the orthogonal mixing matrices $\mathcal{G}_W$ and $\mathcal{G}_Z$ and the masses of the mass
eigenstates are given explicitly in \cite{Albrecht:2009xr}.

Next, the SM fermions are embedded in three possible representations of $SU(2)_L\times SU(2)_R$, that are $(\bm{2},\bm{2}), (\bm{1},\bm{1})$ and
$(\bm{3},\bm{1})\oplus(\bm{1},\bm{3})$. Which fields belong to which multiplets are chosen according to the guidelines provided by
phenomenology. For the realization of the SM quark and lepton sector in the $\text{RS}_c$ model, we refer the reader to ref. \cite{Albrecht:2009xr}.
Moreover, other than SM fields, a number of additional vector-like fermion fields with electric charge $2/3, -1/3$ and $5/3$ are required to fill in the
three representations of the $SU(2)_L\times SU(2)_R$ gauge group. Since we only consider the fermion fields with $(++)$ BCs, we do not discuss the new fermions
which are introduced with $(-+)$ or $(+-)$ choices of the BCs. Furthermore, we will restrict ourselves only to the zero
modes in the KK mode expansion of the fermionic fields with $(++)$ BCs, which are
massless before EWSB and up to small mixing effects with other massive modes after the EWSB, due to the transformation to mass eigenstates,
are identified as the SM quarks and leptons. We have neglected the higher KK fermion modes because their impact
is sub-leading as pointed out previously. The solution of the EOMs of the left and right-handed fermionic zero modes
leads to their bulk profiles, which we denote as $f_{L,R}^{(0)}(y,c_{\Psi})$ and their expressions are given by
\begin{eqnarray}\label{15}
f_L^{(0)}(y,c_{\Psi})=\sqrt{\frac{(1-2c_{\Psi})kL}{e^{(1-2c_{\Psi})kL}-1}}e^{-c_{\Psi}ky},\qquad f_R^{(0)}(y,c_{\Psi})=f_L^{(0)}(y,-c_{\Psi}).
\end{eqnarray}
The bulk mass parameter $c_{\Psi}$ controls the localization of the fermionic zero modes such as for $c_{\Psi}>1/2$, the left-handed fermionic
zero mode is localized towards the UV brane, while for $c_{\Psi}<1/2$, it is localised towards the IR brane. Similarly, from the expression of
the $f_R^{(0)}(y,c_{\Psi})$, the localization of the right-handed fermion zero mode depends on whether $c_{\Psi}<-1/2$ or $c_{\Psi}>-1/2$. For
the SM quarks we will denote the bulk mass parameters $c_Q^i$ for the three left-handed zero mode embedded into bi-doublets of $SU(2)_L\times SU(2)_R$,
while for the right-handed zero mode up and down-type quarks which belong to $(\bm{1},\bm{1})$ and $(\bm{3},\bm{1})\oplus(\bm{1},\bm{3})$ representations, respectively \cite{Carena:2006bn,Albrecht:2009xr}, we assign bulk mass parameters $c_{u,d}^i$, respectively.

The effective 4D Yukawa couplings, relevant for the SM fermion masses and mixings, for the Higgs sector residing on the IR brane are given by \cite{Blanke:2008zb}
\begin{eqnarray}\label{16}
Y_{ij}^{u(d)}=\lambda_{ij}^{u(d)}\frac{e^{kL}}{kL}f_L^{(0)}(y=L,c_Q^i)f_R^{(0)}(y=L,c_{u}^j(c_{d}^j))\equiv\lambda_{ij}^{u(d)}\frac{e^{kL}}{kL}f_i^Qf_j^{u(d)},
\end{eqnarray}
where $\lambda^{u(d)}$ are the fundamental 5D Yukawa coupling matrices. Since the fermion profiles depend exponentially on the bulk
mass parameters, one can recognize from the above relation that the strong hierarchies of quark masses and mixings originate from
the $\mathcal{O}(1)$ bulk mass parameters and anarchic 5D Yukawa couplings $\lambda_{ij}^{u(d)}$. The transformation from the quark flavor eigenbasis
to the mass eigenbasis is performed by means of unitary mixing matrices, which are presented by $\mathcal{U}_{L(R)}$ and $\mathcal{D}_{L(R)}$ for the
up-type left (right) and down-type left (right) quarks, respectively. Moreover, CKM matrix is given by $V_{\text{CKM}}=\mathcal{U}_{L}^{\dagger}\mathcal{D}_{L}$ and the flavor-changing neutral-currents (FCNCs) are induced already at tree level in this model. This happens because the couplings of the fermions with
the gauge bosons involve overlap integrals which contain the profiles of the corresponding fermions and gauge boson leading to non-universal flavor
diagonal couplings. These non-universal flavor diagonal couplings induce off-diagonal entries in the interaction matrix after going to the fermion mass basis,
resulting in tree level FCNCs. These are mediated by the three neutral electroweak gauge bosons $Z$, $Z^{\prime}$ and $Z_H$ as well as by the first KK excitations of the photon and the gluons, although the last one does not contribute to the processes with leptons in the final state. The
expressions of the masses of the SM quarks and the flavor mixing matrices $\mathcal{U}_{L(R)}$, $\mathcal{D}_{L(R)}$ are given explicitly
in terms of the quark profiles and the five-dimensional Yukawa couplings $\lambda_{ij}^{u(d)}$ in \cite{Blanke:2008zb}.

\section{Theoretical Formalism\label{setup}}
The effective weak Hamiltonian for $b\to s \mu^+\mu^-$ transition in the $\text {RS}_c$ model can be written as
\begin{align}\label{18}
H^{\text{RS}_c}_{\text{eff}} = -\frac{4 G_F}{\sqrt{2}} V_{tb} V^*_{ts} \Big[ C^{\text {RS}_c}_7 O_7 &+ C^{\prime \text{RS}_c}_7 O^{\prime}_7 +C^{\text {RS}_c}_9 O_9 + C^{\prime \text {RS}_c}_9 O^{\prime}_9 \notag
\\
&+ C^{\text {RS}_c}_{10} O_{10} + C^{\prime \text {RS}_c}_{10} O^{\prime}_{10} \Big],
\end{align}
where $G_F$ is the Fermi coupling constant and $V_{tb}$,  $V^*_{ts}$ are the elements of the CKM mixing matrix. The involved operators read
\begin{eqnarray}\label{operators}
O_7&=&\frac{e}{16\pi^2}m_b (\bar
s_{L\alpha}\sigma^{\mu\nu}b_{R\alpha})F_{\mu\nu},\notag\\
O_7^{\prime}&=&\frac{e}{16\pi^2}m_b (\bar
s_{R\alpha}\sigma^{\mu\nu}b_{L\alpha})F_{\mu\nu},\notag\\
O_9&=&\frac{e^2}{16\pi^2}(\bar
s_{L\alpha}\gamma^{\mu}b_{L\alpha})\bar \mu\gamma_{\mu} \mu,\notag\\
O_9^{\prime}&=&\frac{e^2}{16\pi^2}(\bar
s_{R\alpha}\gamma^{\mu}b_{R\alpha})\bar \mu\gamma_{\mu} \mu,\notag\\
O_{10}&=&\frac{e^2}{16\pi^2}(\bar
s_{L\alpha}\gamma^{\mu}b_{L\alpha})\bar \mu\gamma_{\mu}\gamma_5 \mu,\notag\\
O_{10}^{\prime}&=&\frac{e^2}{16\pi^2}(\bar
s_{R\alpha}\gamma^{\mu}b_{R\alpha})\bar \mu\gamma_{\mu}\gamma_5 \mu,
\end{eqnarray}
where $e$ is the electromagnetic coupling constant and $m_b$ is the $b$-quark running mass in the $\overline{\text{MS}}$ scheme. In the
$\text {RS}_c$ model the Wilson coefficients in the above effective Hamiltonian can be written as
\begin{equation}\label{19}
C^{(\prime) \text {RS}_c}_i = C^{(\prime) \text{SM}}_i+ \Delta C^{(\prime)}_i,
\end{equation}
where $i=7,9,10$. In the SM case, ignoring tiny contribution, when present, the primed coefficients are zero while the unprimed Wilson coefficients $C_i$ incorporating short distance physics are evaluated through perturbative approach. The factorizable contributions from operators $O_{1-6,8}$ have been absorbed in the effective Wilson coefficients $C^{\text{eff}}_7$ and $C^{\text{eff}}_9$ \cite{Du:2015tda}. The expressions of these effective coefficients involve the functions $h(m_q,q^2)$,
$F_8^{(7,9)}(q^2)$ defined in \cite{Beneke:2001at}, and the functions $F_{1,c}^{(7,9)}(q^2)$, $F_{1,c}^{(7,9)}(q^2)$ given in \cite{Asatryan:2001zw} for low $q^2$
and in \cite{Greub:2008cy} for high $q^2$. The quark masses appearing in these functions are defined in the pole scheme. The long distance non-factorizable contributions of charm loop effects can alter the value of $C^{\text{eff}}_7$ to some extent particularly in the region of charmonium resonances. Modifications $\Delta C_{9,10}^{(\prime)}$, in the $\text {RS}_c$ model, evaluated at the scale $\mathcal{O}(M_{g^{(1)}})$ are given by \cite{Blanke:2008yr}
 \begin{eqnarray}\label{20}
\Delta C_9&=& \frac{\Delta Y_s}{\sin^2\theta_W}-4\Delta
Z_s,\notag\\
\Delta C_{9}^{\prime}&=& \frac{\Delta
Y_s^{\prime}}{\sin^2\theta_W}-4\Delta Z_s^{\prime},\notag\\
\Delta C_{10}&=& -\frac{\Delta Y_s}{\sin^2\theta_W},\notag\\
\Delta C_{10}^{\prime}&=& \frac{\Delta
Y_s^{\prime}}{\sin^2\theta_W},
\end{eqnarray}
where
\begin{eqnarray}\label{21}
\Delta Y_s&=& -\frac{1}{V_{tb}V_{ts}^{\ast }}\sum\limits_X\frac{\Delta_L^{\mu\mu}(X)-\Delta_R^{\mu\mu}(X)}{4M_X^2g_{SM}^2}\Delta_L^{bs}(X),\notag\\
\Delta Y_s^{\prime}&=& -\frac{1}{V_{tb}V_{ts}^{\ast
}}\sum\limits_X\frac{\Delta_L^{\mu\mu}(X)-\Delta_R^{\mu\mu}(X)}{4M_X^2g_{SM}^2}\Delta_R^{bs}(X),\notag\\
\Delta Z_s&=& \frac{1}{V_{tb}V_{ts}^{\ast
}}\sum\limits_X\frac{\Delta_R^{\mu\mu}(X)}{8M_X^2g_{SM}^2\sin^2\theta_W}\Delta_L^{bs}(X),\notag\\
\Delta Z_s^{\prime}&=& \frac{1}{V_{tb}V_{ts}^{\ast
}}\sum\limits_X\frac{\Delta_R^{\mu\mu}(X)}{8M_X^2g_{SM}^2\sin^2\theta_W}\Delta_R^{bs}(X).
\end{eqnarray}
The sums run over the neutral gauge bosons $X=Z, Z^{\prime}, Z_{H}$ and $A^{(1)}$ with
$g_{SM}^2=\frac{G_F}{\sqrt 2}\frac{\alpha}{2\pi\sin^2\theta_{W}}$. $\Delta C^{(\prime)}_9$ and $\Delta C^{(\prime)}_{10}$ evaluated at the scale $M_{g^{(1)}}$ do not need to be evolved to $\mu_b$ scale. In the case of $\Delta C_7^{(\prime)}$, detailed calculation with the set of assumptions consistent with the calculations of
$\Delta C_{9,10}^{(\prime)}$ is given in Appendix C of Ref. \cite{Biancofiore:2014wpa}, where $\Delta C_7$ and $\Delta C_7^{\prime}$ are evaluated at the $M_{g^{(1)}}$ scale. The evolution at the scale $\mu_b$ is given by the following master formula \cite{Blanke:2012tv}
\begin{eqnarray}\label{22}
\Delta C^{(\prime)}_7 (\mu_b) = 0.429 \Delta C^{(\prime)}_7 (M_{g^{(1)}}) +0.128 \Delta C^{(\prime)}_8 (M_{g^{(1)}}).
\end{eqnarray}
The decay amplitude for $\Lambda_b \to \Lambda \mu^+ \mu^-$ can be obtained by sandwiching the
effective Hamiltonian displayed in Eq. (\ref{18}) within the baryonic states
\begin{eqnarray}\label{23}
\mathcal{M}_{\text {RS}_c} (\Lambda_b \to \Lambda \mu^+ \mu^-) = \frac{G_{F}\alpha}{\sqrt{2}\pi}V_{tb}V_{ts}^{\ast}\bigg[\langle\Lambda(k)|\bar s\gamma_{\mu}(C_{9}^{\text {RS}_c} P_L+C_{9}^{\prime \text {RS}_c} P_R)b|\Lambda_{b}(p)\rangle
(\bar\mu\gamma^{\mu}\mu)\notag\\
+[\langle\Lambda(k)|\bar s\gamma_{\mu}(C_{10}^{\text {RS}_c} P_L+C_{10}^{\prime \text {RS}_c} P_R)b|\Lambda_{b}(p)\rangle
(\bar\mu\gamma^{\mu} \gamma^5 \mu)\notag\\
-\frac{2m_{b}}{q^{2}}\langle\Lambda(k)|\bar s i\sigma_{\mu\nu}q^{\nu}(C_{7}^{\text {RS}_c} P_R + C_{7}^{ \prime \text {RS}_c} P_L) b| \Lambda_{b}(p) \rangle\bar\mu
\gamma^{\mu}\mu \bigg].
\end{eqnarray}
The matrix elements involved in the expression of decay amplitude are given in \cite{Boer:2014kda} written in helicity basis in terms of FF. The detailed calculation of FFs in lattice QCD is carried out in \cite{Detmold:2016pkz}, which will be used in our numerical analysis.
The angular decay distribution of the four-fold decay $\Lambda_{b} \to \Lambda(\to p\pi)\mu^{+}\mu^{-}$, with an unpolarized $\Lambda_{b}$, can be written as \cite{Gutsche:2013pp,Boer:2014kda}
\begin{eqnarray}\label{24}
\frac{d^4\Gamma}{ds \ d\cos{\theta_{\Lambda}} \ d\cos {\theta}_l \ d\phi} &=& \frac{3}{8 \pi} \bigg[K_{1ss} \sin^2{\theta_l}+K_{1cc} \cos^2{\theta_l}+ K_{1c} \cos{\theta_l}  \notag \\
&+& (K_{2ss} \sin^2{\theta_l}+K_{2cc} \cos^2{\theta_l}+ K_{2c} \cos{\theta_l}) \cos{\theta_{\Lambda}} \notag \\
&+&(K_{3sc} \sin{\theta_l} \cos{\theta_l}+K_{3s} \sin{\theta_l}) \sin{\theta_{\Lambda}} \sin{\phi} \notag \\
&+&(K_{4sc} \sin{\theta_l} \cos{\theta_l}+K_{4s} \sin{\theta_l}) \sin{\theta_{\Lambda}} \cos{\phi}  \bigg],
\end{eqnarray}
where $K$'s represent the angular coefficients which are functions of $s=q^2$. Here we concentrate on the observables
which have been measured experimentally so that we compare our analysis with
experimental data. For the decay under consideration decay rate and longitudinal polarization of the daughter baryon $\Lambda$ are
\begin{equation}\label{25}
\frac{d\Gamma}{ds}=2K_{1ss}+K_{1cc} , \qquad\quad
F_{L} =\frac{%
	2K_{1ss}-K_{1cc}}{2K_{1ss}+K_{1cc}}.
\end{equation}%
Forward-backward asymmetry with respect to leptonic and baryonic angles is given as
\begin{equation}\label{AFB}
A_{FB}^{l}=\frac{3K_{1c}}{4K_{1ss}+2K_{1cc}}, \qquad\quad
A_{FB}^{\Lambda }=\frac{2K_{2ss}+K_{2cc}}{4K_{1ss}+2K_{1cc}}.
\end{equation}%
The combined FB asymmetry is
\begin{equation}\label{26}
A_{FB}^{l\Lambda }=\frac{3K_{2c}}{8K_{1ss}+4K_{1cc}}.
\end{equation}%
The uncertainties in the decay rate are larger as it strongly depends on hadronic Form Factors. The other observables being ratio of angular coefficients, are more sensitive to NP effects but less sensitive to hadronic FFs.

\section{Constraints and generation of the parameter space of the $\text{RS}_c$ model\label{Rscbounds}}

In this section we consider the relevant constraints on
the parameter space of the $\text{RS}_c$ model coming from the direct searches at the
LHC \cite{Aad:2015fna,Sirunyan:2017uhk}, EW precision tests \cite{Malm:2013jia,Malm:2014gha}, the latest measurements of the
Higgs signal strengths at the LHC \cite{Malm:2014gha} and from $\Delta F=2$ flavor observables \cite{Blanke:2008zb}.

Starting with the direct searches, current measurements at the LHC for resonances decaying to $t\bar t$ pair constrain the
lightest KK gluon mass $M_{g^{(1)}}>3.3$ TeV at $95\%$ confidence level \cite{Sirunyan:2017uhk}. Further, in the $\text{RS}_c$ model, EW precision measurements permit to have masses of the lowest KK gauge bosons in the few TeV range.
For example, a tree-level analysis of the S and T parameters leads to
$M_{g^{(1)}}>4.8$ TeV for the lightest KK gluon and KK photon masses \cite{Malm:2013jia}.
Furthermore, a comparison of the predictions of all relevant Higgs decays in the $\text{RS}_c$ model
with the latest data from the LHC shows that the signal rates for $pp\rightarrow h\rightarrow
ZZ^{\ast}, WW^{\ast}$ provide the most stringent bounds, such that KK gluon masses lighter
than $22.7$ $\text{TeV}\times(y_{\star}/3)$ in the brane-Higgs case
and $13.2$ $\text{TeV}\times(y_{\star}/3)$ in the narrow bulk-Higgs
scenario are excluded at $95\%$ probability \cite{Malm:2014gha}, where
$y_{\star}=\mathcal{O}(1)$ free parameter is defined as the
upper bound on the anarchic 5D Yukawa couplings such that $|\lambda_{ij}^{u(d)}|\le
y_{\star}$. This implies that $y_{\star}=3$ value, coming from the perturbativity bound of the RS model, will lead to much stronger
bounds from Higgs physics than those emerging from the EW precision tests.
In general, one can lower these bounds by considering smaller values of $y_{\star}$.
However one should keep in mind that lowering the bounds upto KK gauge bosons
masses implied by EW precision constraints, $M_{g^{(1)}}=4.8$ TeV, will require
too-small Yukawa couplings, $y_{\star}<0.3$ for the
brane-Higgs scenario \cite{Malm:2014gha}, which will reinforce the RS flavor problem
because of enhanced corrections to $\epsilon_K$. Therefore, moderate bounds
on the value of the $y_{\star}$ should be considered by relatively increasing the KK scale,
in order to avoid constraints from both flavour observables and Higgs physics.

Next, in analogy to our previous analysis \cite{Lu:2016pfs}, we explore the parameter space of the $\text {RS}_c$ model
by generating two sets of anarchic 5D Yukawa matrices, whose entries satisfy $|\lambda_{ij}^{u(d)}|\le
y_{\star}$ with $y_{\star}=1.5$ and $3$. Further, we choose the nine quark bulk-mass parameters $c_{Q,u,d}$,
which together with the 5D Yukawa matrices reproduce the correct values of the quark masses evaluated at the scale $\mu=3$ TeV, CKM mixing
angles and the Jarlskog determinant, all within their respective $2\sigma$ ranges. For muon, we take $c_{\mu}=0.7$ as lepton flavor-conserving
couplings are found to be almost independent of the chosen value as far as $c_l>0.5$ \cite{Blanke:2008yr}. Additionally, from the $\Delta F=2$ flavor observables, we apply the constraints from $\epsilon_K$, $\Delta M_K$ and $\Delta M_{B_s}$ observables, where we set the required input parameters, as given in Table 2 of \cite{Lu:2016pfs}, to their central values and allow the resulting observables to deviate by $\pm 30\%$, $\pm 50\%$ and $\pm 30\%$, respectively in analogy to the analysis \cite{Blanke:2008zb}. For further details on the parameter scan, we refer
the reader to \cite{Lu:2016pfs,Blanke:2008zb}.


\section{Numerical Analysis\label{num}}
\subsection{Wilson coefficients}
\begin{figure*}[ht]
\begin{center}
\includegraphics[scale=0.40]{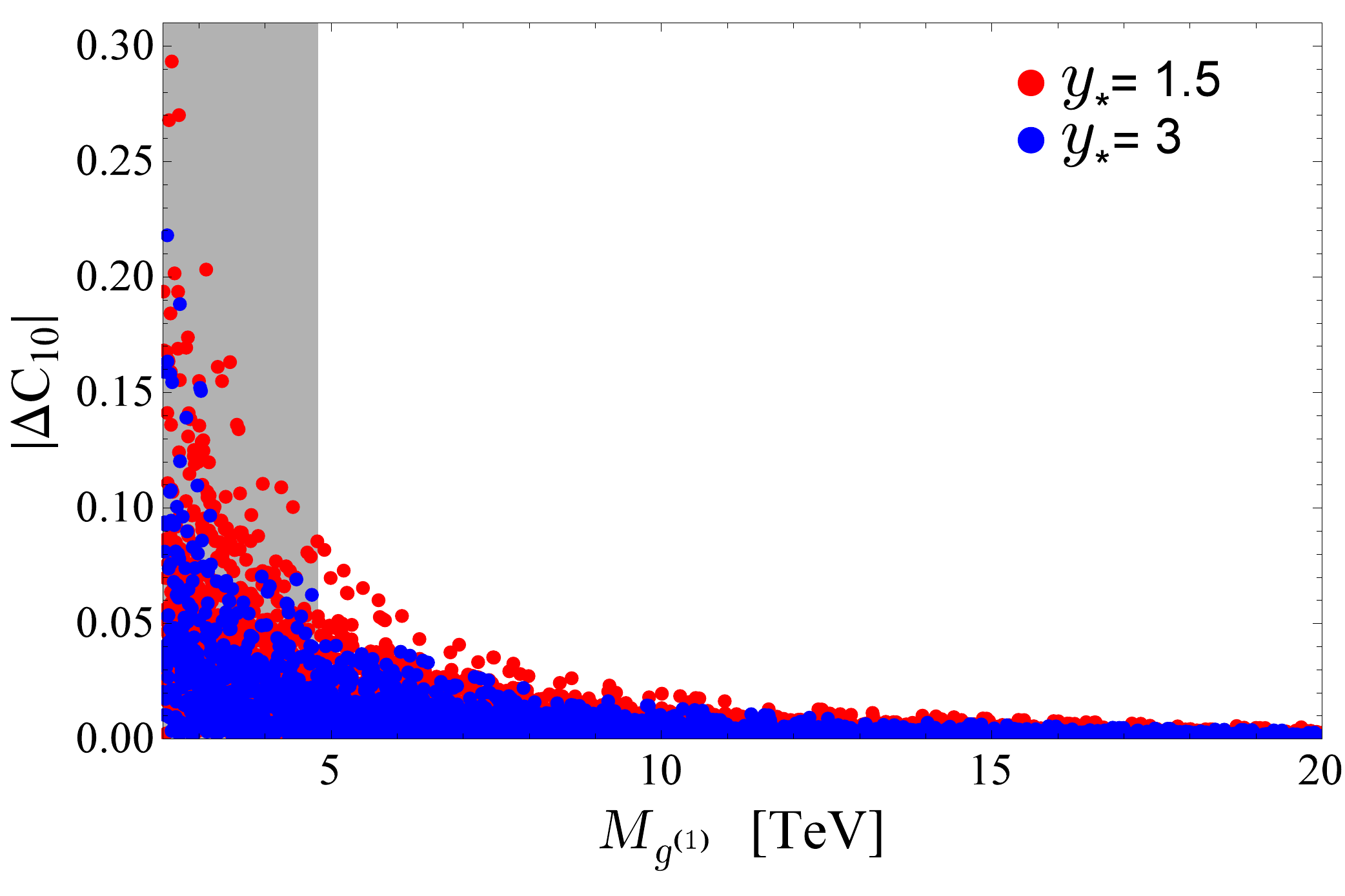}
\end{center}
\caption{(color online) The $\text{RS}_c$ contribution to $|\Delta C_{10}|$ as a function of the KK gluon mass $M_{g^{(1)}}$ for two different values of $y_{\star}$. The gray region is excluded by the analysis of electroweak precision measurements.}
\label{Figure1}
\end{figure*}

The generated 5D parameter points consisting of Yukawa coupling matrices and bulk mass parameters, fulfilling all the relevant
constraints, are used to evaluate the Wilson coefficients in the $\text {RS}_c$ model. In Fig. \ref{Figure1}, we show the
dependence of $|\Delta C_{10}|$ Wilson coefficient on the mass of lowest KK gluon $M_{g^{(1)}}$ taken in the range $2.45$ to 20 TeV.
The red and blue scatter points represent the cases of $y_{\star}=1.5$ and $3$, respectively. The gray region is excluded by the analysis
of EW precision observables. It is clear that the smaller values of $M_{g^{(1)}}$ give larger deviations. Moreover, for a fixed value of
$M_{g^{(1)}}$ a range of predictions for possible deviations are present for both cases of $y_{\star}$ such that the maximum allowed
deviation for $|\Delta C_{10}|$ in the case of $y_{\star}=1.5$ are generally greater than the case of $y_{\star}=3$. This is due to the fact that in the case of $y_{\star}=3$, the SM fermions are more elementary as their profiles are localized towards the UV brane to a greater extent
compared to the $y_{\star}=1.5$ case leading to more suppressed FCNC and subsequently smaller deviations in comparison to the case of $y_{\star}=1.5$.
\begin{figure*}[ht!]
\centering
\begin{scriptsize}
\begin{tabular}{cc}
\includegraphics[scale=0.35]{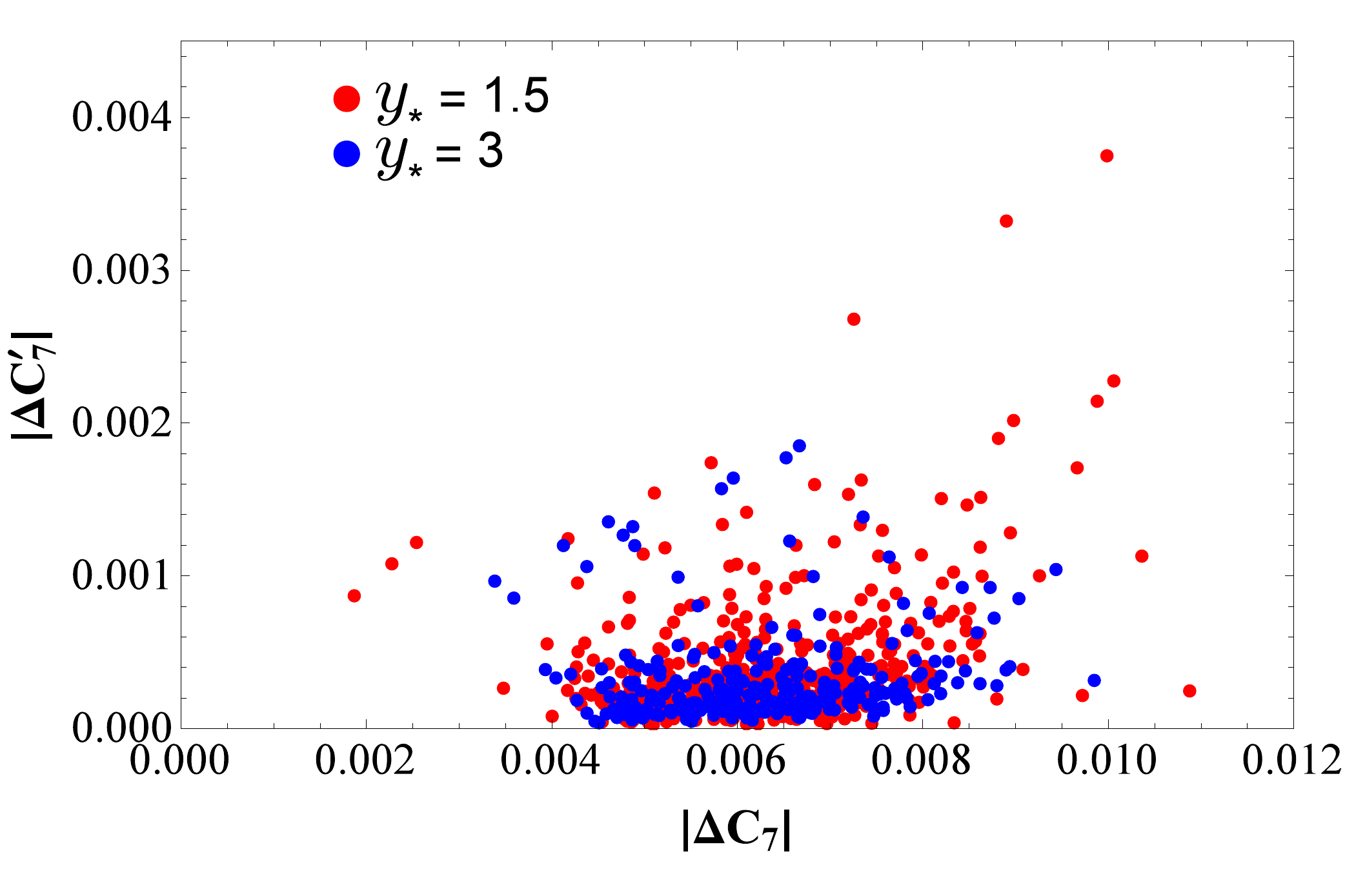} \ \ \
& \ \ \ \includegraphics[scale=0.35]{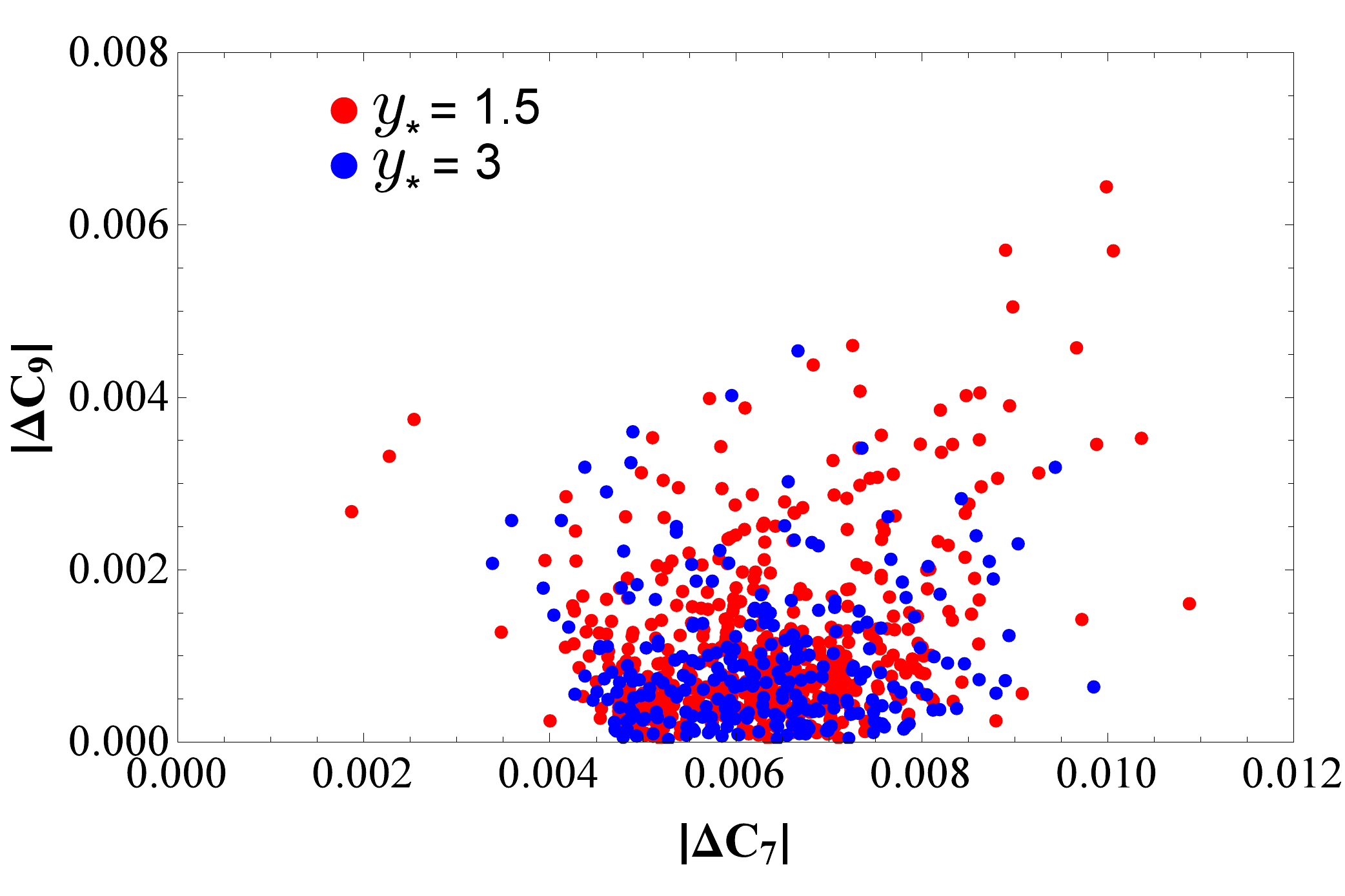}\\
\hspace{0.4cm}(a)&\hspace{1.0cm}(b)\\
\includegraphics[scale=0.35]{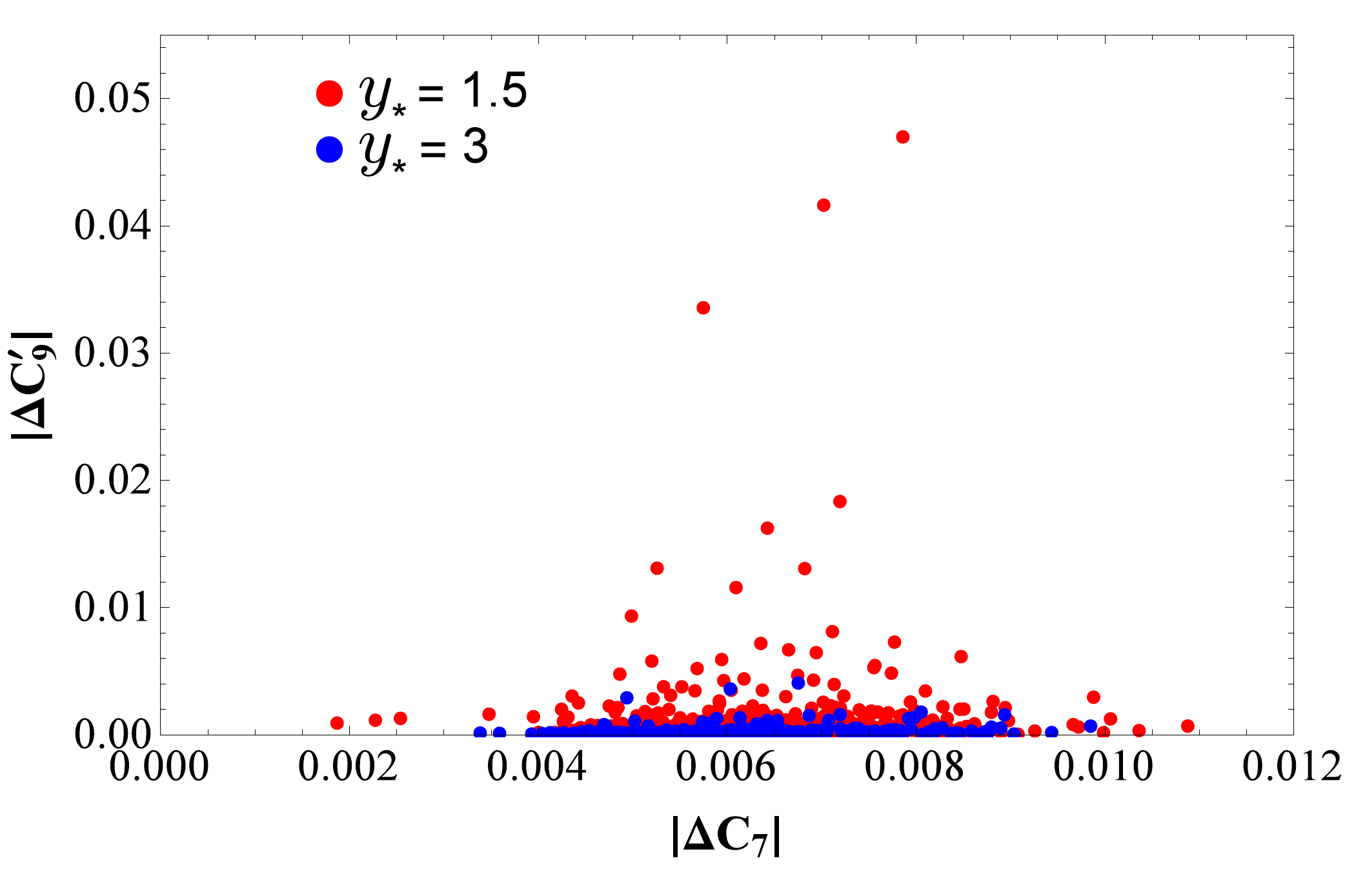} \ \ \
& \ \ \ \includegraphics[scale=0.35]{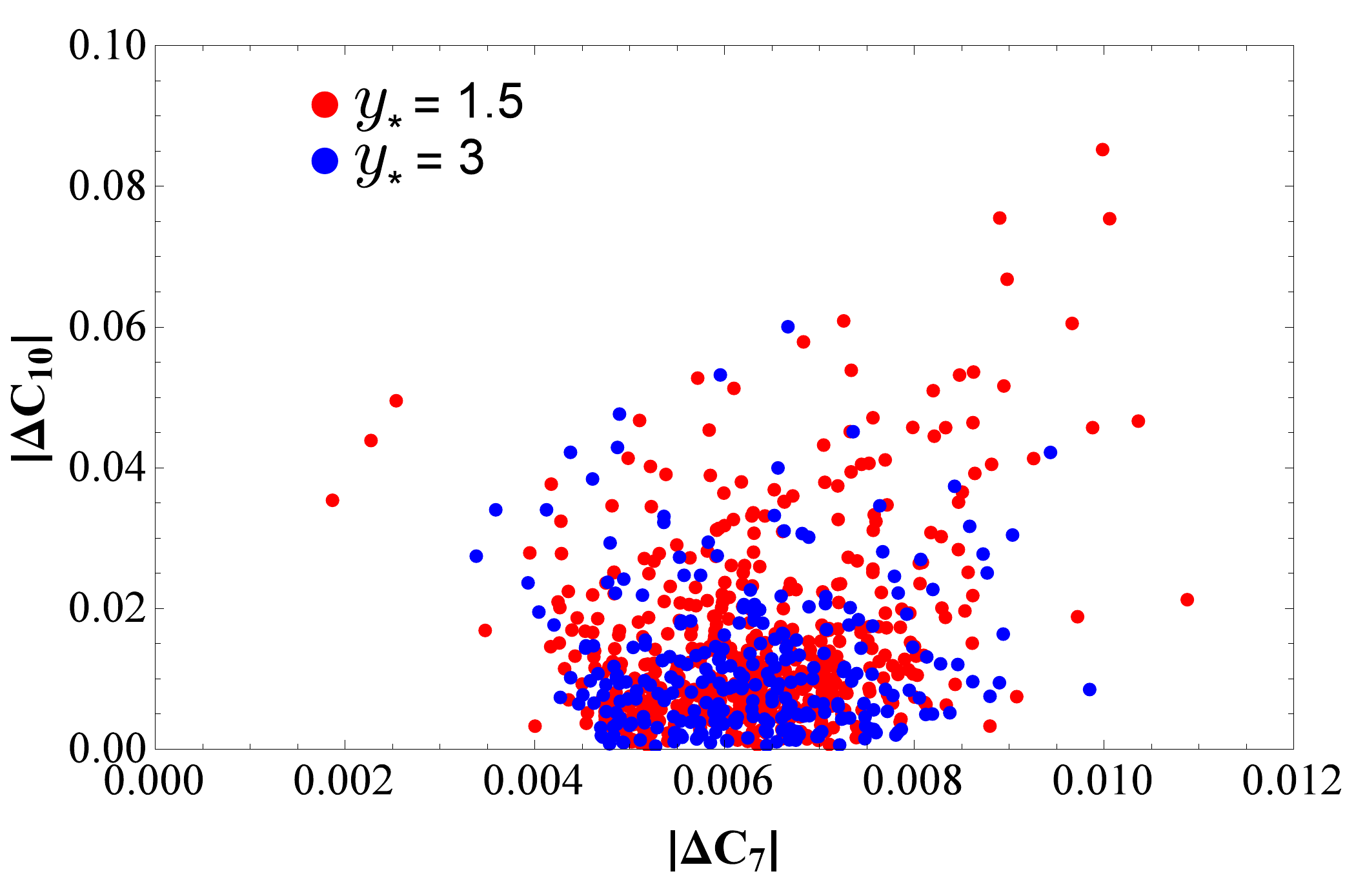}\\
\hspace{0.3cm}(c)&\hspace{1.0cm}(d)\\
\includegraphics[scale=0.35]{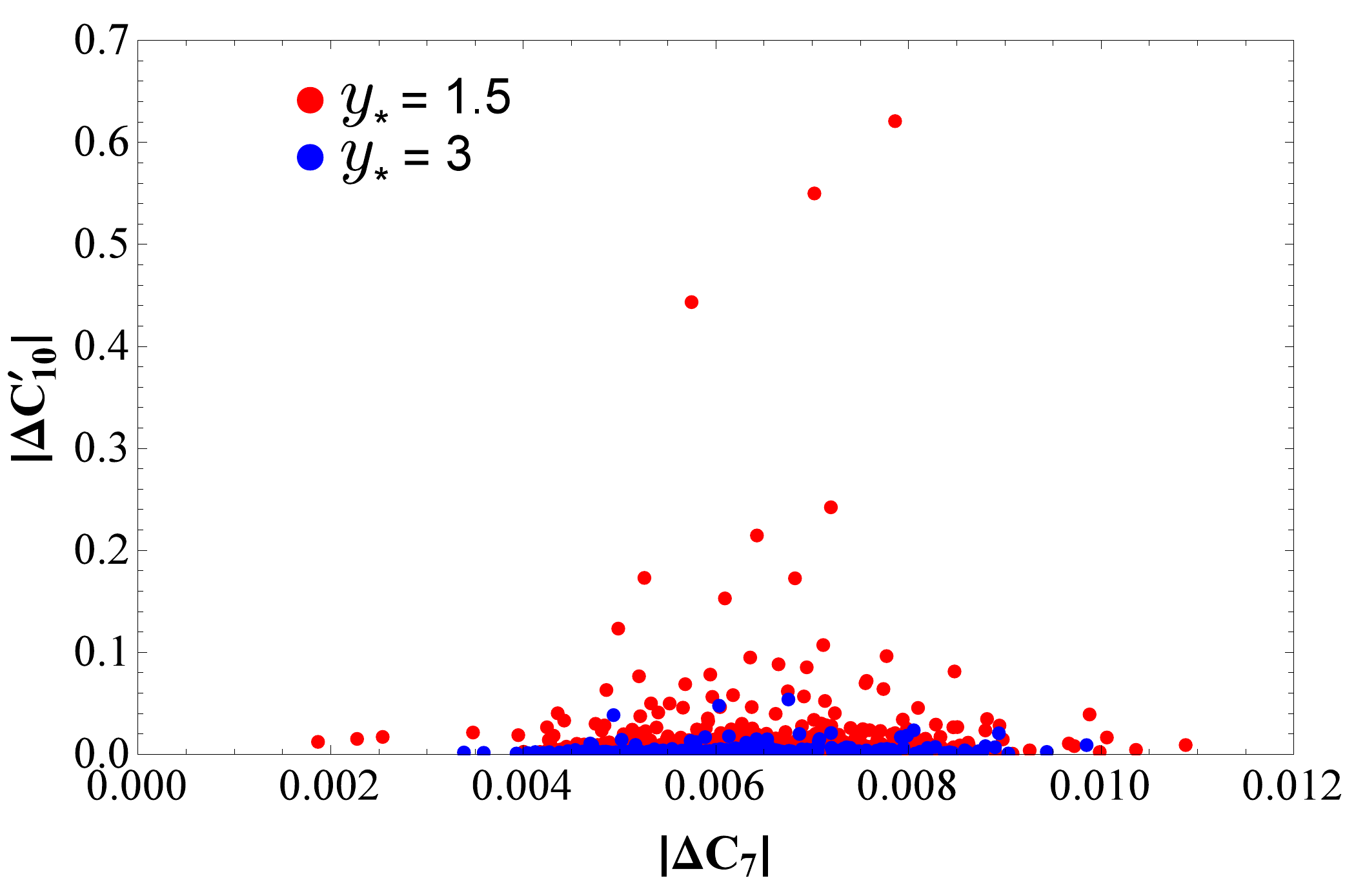} \ \ \
& \ \ \ \includegraphics[scale=0.345]{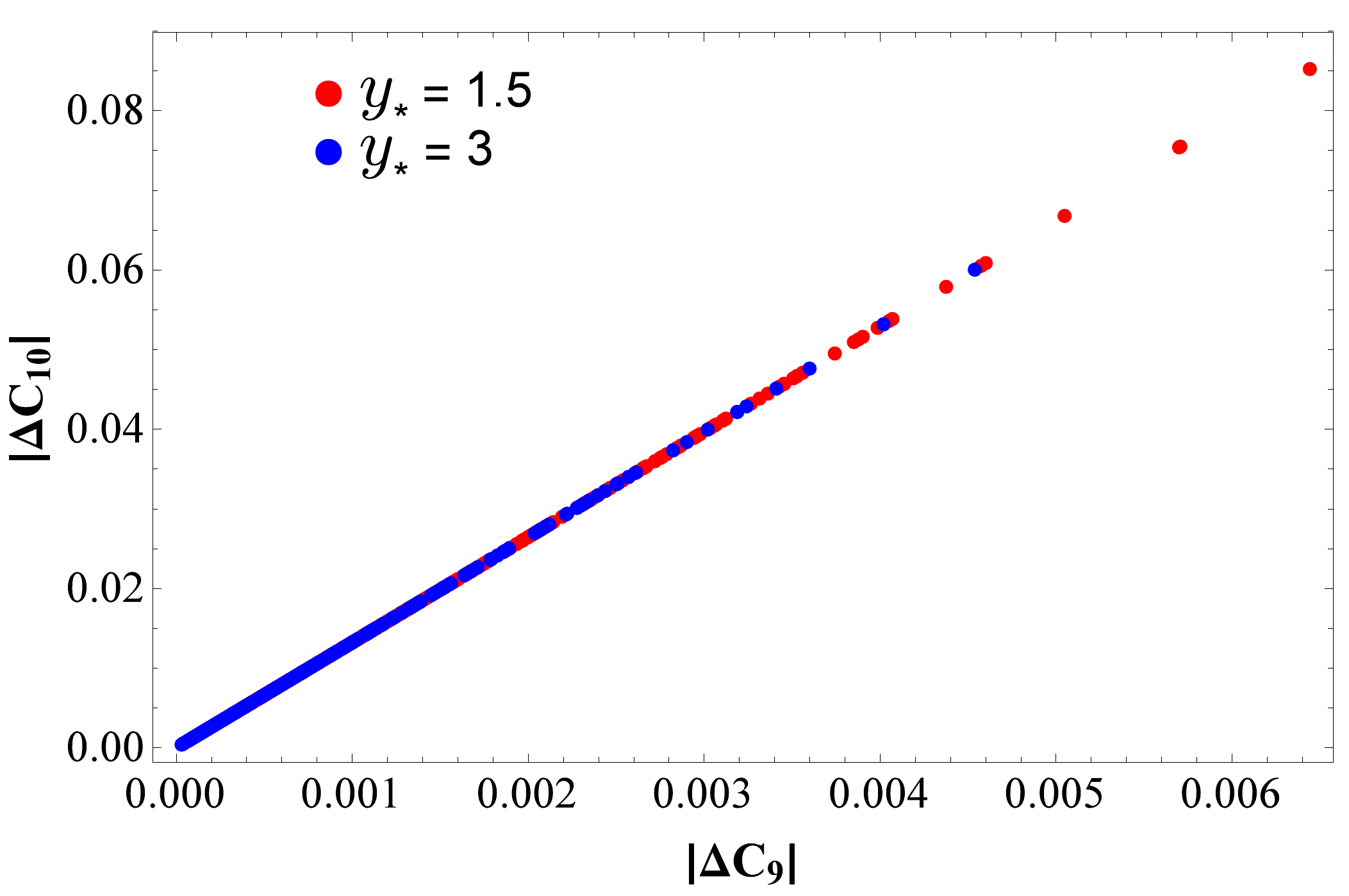}\\
\hspace{0.3cm}(e)&\hspace{1.0cm}(f)\\
\end{tabular}
\end{scriptsize}
\caption{(color online) Correlations plots between the Wilson coefficients $|\Delta C^{(\prime)}_{7,\; 9,\; 10}|$ of the $\text{RS}_c$ model for a fixed value of
$M_{g^{(1)}}=4.8$ TeV. The coefficients $\Delta C^{(\prime)}_7$ are calculated at the $\mu_b$ scale. The red and blue points correspond to $y_{\star}=1.5$ and $3$, respectively.\label{coplots}}
\end{figure*}
Observing the fact that the deviations for all $|\Delta C_{i}^{(\prime)}|$
for $M_{g^{(1)}}>10$ TeV are so small, as clear from Fig. \ref{Figure1} in the case of $|\Delta C_{10}|$, that the observables will almost remain unaffected,
we limit the range for $M_{g^{(1)}}$ from $4.8$ TeV to $10$ TeV, where the lower value is implied by the EW precision constraints. As we are interested in the largest possible deviations of $|\Delta C_{i}^{(\prime)}|$, for a given allowed value of $M_{g^{(1)}}$, so we will take the $y_{\star}=1.5$ case and by considering five different values of $M_{g^{(1)}}\in[4.8,10]$, we obtain the maximum possible deviation of each Wilson coefficient. The resultant values
will be used for evaluating the effects on the angular observables of interest for each considered value of $M_{g^{(1)}}$ in next section i.e., Sec. \ref{Angul-Obs}.

In Fig. \ref{coplots}, we show the correlation plots between $|\Delta C^{(\prime)}_{7,\; 9,\; 10}|$ obtained for the fixed value of $M_{g^{(1)}}=4.8$ TeV. The
maximum possible deviations from the SM values in this case are
\begin{eqnarray}
\left| \Delta C_7\right|_{max} = 0.011, \text{ \ \ \ \ } \left|  \Delta C_9 \right|_{max} = 0.0064,  \text{ \ \ \ } \left|  \Delta C_{10} \right|_{max} = 0.085, \notag \\
\left| \Delta C_{7}^{\prime} \right|_{max}=0.0037, \text{ \ \ \ \ } \left| \Delta C_{9}^{\prime} \right|_{max}= 0.047, \text{ \ \ \  } \left| \Delta C_{10}^{\prime}\right|_{max} = 0.621. \notag
\end{eqnarray}
It is found that $|\Delta C_{9}|$ and $|\Delta C_{10}|$ are linearly correlated, as shown in Fig. \ref{coplots}(f), and same is true for
each pair $|\Delta C_{i}^{(\prime)}|$ with $i=9,10$.

\subsection{Angular observables\label{Angul-Obs}}
\begin{table}
\centering \caption{Default values of the input parameters used in the calculations \cite{Patrignani:2016xqp,Detmold:2016pkz}.}\label{table1}
\begin{tabular}{lll}
\hline\hline
$G_{F}=1.16638\times 10^{-5}$ GeV$^{-2}$  & $m_t^{\text{pole}}=174.2\pm1.4$ GeV & $m_{\pi}=0.135$ GeV  \\
$\alpha_s(m_Z)= 0.1182\pm0.0012$  & $m_b^{\text{pole}}=4.78\pm0.06$ GeV  & $m_K=0.494$ GeV   \\
$\alpha(\mu_b)=1/133.28$ & $m_c^{\text{pole}}=1.67\pm0.07$ GeV &  $m_B=5.279$ GeV    \\
$m_{W}=80.385\pm0.015$ GeV & $m_b=4.18_{-0.03}^{+0.04}$ GeV &  $m_{\Lambda_b}=5.619$ GeV   \\
$m_{Z}=91.1876\pm0.0021$ GeV &  $m_c=1.27\pm0.03$ GeV& $\tau_{\Lambda_b}=(1.466\pm0.010)$ ps   \\
$|V_{tb}V_{ts}^{\ast}|=0.04152$     &  $m_s=0.096_{-0.004}^{+0.008}$ GeV   &  $m_{\Lambda}=1.116$ GeV     \\
$\alpha_{\Lambda}=0.642\pm0.013$   &  $\mu_b=4.2$ GeV      &      \\
\hline\hline
\end{tabular}
\end{table}
In this section we discuss the numerical results computed for different angular observables both in the SM and for the $\text{RS}_c$ model.
The input parameters used in the calculations are included in Table \ref{table1}. The presented results include the uncertainty in the hadronic FFs, which are non-perturbative quantities. For this, we utilize the lattice QCD calculations \cite{Detmold:2016pkz}, both in the low and high $q^2$ ranges, which till todate are considered as most accurate in the literature. To improve the accuracy, we have used the numerical values for the short-distance Wilson coefficients, with NNLL accuracy, at the low energy scale $\mu_b=4.2$ GeV, given in Table \ref{table2}.
\begin{table}
\centering \caption{The SM Wilson coefficients up-to NNLL accuracy given at $\mu_b=4.2$ GeV scale.}\label{table2}
\begin{tabular}{llllll}
\hline\hline
$C_1=-0.294$   & $C_2=1.017$  & $C_3=-0.0059$ & $C_4=-0.087$ & $C_5=0.0004$ \\
$C_6=0.0011$ & $C_7=-0.324$ & $C_8=-0.176$ & $C_9=4.114$& $C_{10}=-4.193$ \\
\hline\hline
\end{tabular}
\end{table}

The numerical results for the angular observables in appropriate bins are
shown in Tables \ref{low-s-bin-values} and \ref{high-s-bin-values}, where a comparison is presented between the predictions obtained for five different values of $M_{g^{(1)}}$ in the $\text{RS}_c$ model
(for $y_{\star}=1.5$ ) to that of the SM estimates and with the experimental measurements, where available.
The whole spectrum of di-muon mass squared (s $ \epsilon $ $ \{ s_{min}=4m^2_{\mu}, s_{max}=( m^2_{\Lambda_b} -m^2_{\Lambda}) \}$) has not been discussed as the region $s \;  \in $ $ [8,15 ] $ GeV$^2$ is expected to receive sizable corrections from charmonium loops that violate quark-hadron duality. Hence the regions $s \;  \in $ $ [0.1 ,8] $ GeV$^2$ and $s\;  \in   [15,20 ] $ GeV$^2$ have been considered in order to avoid the long distance effects of charmonium resonances arising when lepton pair momenta approaches the masses of $J/\psi$ family. It can be seen that the results in the $\text{RS}_c$ model for most of the observables show little deviation from the SM predictions. Maximum deviation from the SM results has been observed for $M_{g^{(1)}}=4.8$ TeV and the difference gradually decreases as one moves from $M_{g^{(1)}}=4.8$ TeV to $M_{g^{(1)}}=10$ TeV.
\begin{table*}[th]
	\caption{Numerical results of the observables (low $s$ region) in the $\Lambda_b \rightarrow \Lambda (\rightarrow p \pi) \mu^+ \mu^-$ decay,
 obtained for the SM and the $\text{RS}_c$ model with $y_{\star} =1.5$ case, in different bins of low $s$. Experimentally measured values are taken from \cite{Aaij:2015xza}.} \label{low-s-bin-values}
\begin{scriptsize}
	\begin{tabular}{l||llllll}
		\hline\hline
		& $\ \left\langle \frac{d\mathcal{B} }{ds}\times 10^{-7}\right\rangle $ & $\ \ \ \
		\ \ \left\langle F_{L}\right\rangle $ & $\ \ \ \ \ \ \left\langle
		A_{FB}^{\ell }\right\rangle $ & $\ \ \ \ \  \left\langle
		A_{FB}^{\Lambda}\right\rangle $ & $\ \ \ \ \ \ \left\langle A_{FB}^{l\Lambda
		}\right\rangle $ \\ \hline\hline
		$\lbrack 0.1,2]%
		\begin{tabular}{l}
		$\text{SM}$ \\
		$\text{RS}_c|_{M_{g^{(1)}}}=4.8$ \\
		$\text{RS}_c|_{M_{g^{(1)}}}=6.1$ \\
		$\text{RS}_c|_{M_{g^{(1)}}}=7.4$ \\
		$\text{RS}_c|_{M_{g^{(1)}}}=8.7$ \\
		$\text{RS}_c|_{M_{g^{(1)}}}=10$ \\
		$ \text{LHCb} $
		\end{tabular}%
		$ & $%
		\begin{tabular}{l}
		$ \ 0.238_{-0.230}^{+0.230}$ \\
		$ \ 0.219_{-0.217}^{+0.218}$ \\
		$ \ 0.225_{-0.217}^{+0.219}$ \\
		$ \ 0.229_{-0.222}^{+0.224}$ \\
		$ \ 0.232_{-0.223}^{+0.224}$ \\
		$ \ 0.233_{-0.225}^{+0.228}$ \\
		$ \ 0.36_{-0.112}^{+0.122}$
		\end{tabular}%
		$ & $%
		\begin{tabular}{l}
		$ \ \ \ \ 0.535_{-0.078}^{+0.065}$ \\
		$ \ \ \ \ 0.552_{-0.084}^{+0.069}$\\
		$ \ \ \ \ 0.545_{-0.082}^{+0.067}$ \\
		$ \ \ \ \ 0.542_{-0.081}^{+0.067}$ \\
		$ \ \ \ \ 0.540_{-0.080}^{+0.066}$ \\
		$ \ \ \ \ 0.539_{-0.080}^{+0.066}$ \\
		$ \ \ \ \ 0.56_{-0.566}^{+0.244}$
		\end{tabular}%
		$ & $%
		\begin{tabular}{l}
		$ \ \ \ \ 0.097_{-0.007}^{+0.006}$ \\
		$ \ \ \ \ 0.093_{-0.006}^{+0.005}$ \\
		$ \ \ \ \ 0.095_{-0.006}^{+0.005}$ \\
		$ \ \ \ \ 0.095_{-0.007}^{+0.006}$ \\
		$ \ \ \ \ 0.096_{-0.007}^{+0.006}$ \\
		$ \ \ \ \ 0.096_{-0.007}^{+0.006}$ \\
		$ \ \ \ \ 0.37_{-0.481}^{+0.371}$
		\end{tabular}%
		$ & $%
		\begin{tabular}{l}
		$ \ \  -0.310_{-0.008}^{+0.015}$ \\
		$ \ \  -0.313_{-0.004}^{+0.013}$ \\
		$ \ \ -0.313_{-0.006}^{+0.014}$ \\
		$ \ \  -0.312_{-0.007}^{+0.015}$ \\
		$ \ \  -0.312_{-0.007}^{+0.015}$ \\
		$ \ \ -0.311_{-0.007}^{+0.015}$ \\
		$ \ \ -0.12_{-0.318}^{+0.344}$
		\end{tabular}%
		$ & $%
		\begin{tabular}{l}
		$ \ \ -0.031_{-0.002}^{+0.003}$ \\
		$ \ \ -0.030_{-0.002}^{+0.003}$ \\
		$ \ \ -0.030_{-0.002}^{+0.003}$ \\
		$ \ \ -0.030_{-0.002}^{+0.003}$ \\
		$ \ \ -0.031_{-0.002}^{+0.003}$ \\
		$ \ \ -0.031_{-0.002}^{+0.003}$  \\
		$ \ \ \ \ \ \ \ \ - $
		\end{tabular}
		$ \\ \hline
		$\lbrack 2,4]%
		\begin{tabular}{l}
		$\text{SM}$ \\
		$\text{RS}_c|_{M_{g^{(1)}}}=4.8$ \\
		$\text{RS}_c|_{M_{g^{(1)}}}=6.1$ \\
		$\text{RS}_c|_{M_{g^{(1)}}}=7.4$ \\
		$\text{RS}_c|_{M_{g^{(1)}}}=8.7$ \\
		$\text{RS}_c|_{M_{g^{(1)}}}=10$ \\
		$ \text{LHCb} $
		\end{tabular}%
		$ & $%
		\begin{tabular}{l}
		$ \ 0.180_{-0.123}^{+0.123}$ \\
		$ \ 0.171_{-0.117}^{+0.118}$ \\
		$ \ 0.173_{-0.117}^{+0.118}$ \\
		$ \ 0.175_{-0.118}^{+0.119}$ \\
		$ \ 0.176_{-0.119}^{+0.119}$ \\
		$ \ 0.177_{-0.120}^{+0.120}$ \\
		$ \ 0.11_{-0.091}^{+0.120}$
		\end{tabular}%
		$ & $%
		\begin{tabular}{l}
		$ \ \ \ \ 0.855_{-0.012}^{+0.008}$ \\
		$ \ \ \ \ 0.860_{-0.006}^{+0.008}$  \\
		$ \ \ \ \ 0.859_{-0.008}^{+0.008}$ \\
		$ \ \ \ \ 0.858_{-0.009}^{+0.008}$ \\
		$ \ \ \ \ 0.857_{-0.010}^{+0.008}$ \\
		$ \ \ \ \ 0.857_{-0.011}^{+0.008}$ \\
		$ \ \ \ \ \ \ \ - $
		\end{tabular}%
		$ & $%
		\begin{tabular}{l}
		$ \ \ \ \ 0.054_{-0.030}^{+0.037}$ \\
		$ \ \ \ \ 0.040_{-0.026}^{+0.035}$ \\
		$ \ \ \ \ 0.045_{-0.028}^{+0.036}$ \\
		$ \ \ \ \  0.048_{-0.028}^{+0.036}$ \\
		$ \ \ \ \ 0.050_{-0.029}^{+0.036}$ \\
		$ \ \ \ \ 0.051_{-0.030}^{+0.037}$ \\
		$ \ \ \ \ \ \  - $
		\end{tabular}%
		$ & $%
		\begin{tabular}{l}
		$ \ \  -0.306_{-0.012}^{+0.022}$ \\
		$ \ \ -0.311_{-0.005}^{+0.016}$ \\
		$ \ \ -0.311_{-0.008}^{+0.018}$ \\
		$ \ \ -0.310_{-0.009}^{+0.020}$ \\
		$ \ \ -0.310_{-0.010}^{+0.020}$ \\
		$ \ \ -0.309_{-0.010}^{+0.021}$ \\
		$ \ \ \ \ \ \ \ \ - $
		\end{tabular}%
		$ & $%
		\begin{tabular}{l}
		$ \ \ -0.016_{-0.009}^{+0.008}$ \\
		$ \ \ -0.013_{-0.010}^{+0.009}$ \\
		$ \ \ -0.014_{-0.009}^{+0.008}$ \\
		$ \ \ 0.015_{-0.009}^{+0.008}$ \\
		$ \ \ -0.015_{-0.009}^{+0.008}$ \\
		$ \ \ -0.016_{-0.009}^{+0.008}$ \\
		$ \ \ \ \ \ \ \ \ - $
		\end{tabular}
		$ \\ \hline
		$\lbrack 4,6]%
		\begin{tabular}{l}
		$\text{SM}$ \\
		$\text{RS}_c|_{M_{g^{(1)}}}=4.8$ \\
		$\text{RS}_c|_{M_{g^{(1)}}}=6.1$ \\
		$\text{RS}_c|_{M_{g^{(1)}}}=7.4$ \\
		$\text{RS}_c|_{M_{g^{(1)}}}=8.7$ \\
		$\text{RS}_c|_{M_{g^{(1)}}}=10$ \\
		$  \text{LHCb} $
		\end{tabular}%
		$ & $%
		\begin{tabular}{l}
		$ \  0.232_{-0.110}^{+0.110}$ \\
		$ \   0.224_{-0.108}^{+0.108}$ \\
		$ \  0.227_{-0.108}^{+0.109}$ \\
		$ \   0.228_{-0.109}^{+0.109}$ \\
		$ \  0.229_{-0.109}^{+0.109}$ \\
		$ \  0.230_{-0.110}^{+0.110}$ \\
		$ \ 0.02_{-0.010}^{+0.091}$
		\end{tabular}%
		$ & $%
		\begin{tabular}{l}
		$ \ \ \ \ 0.807_{-0.012}^{+0.018}$ \\
		$ \ \ \ \ 0.806_{-0.016}^{+0.021}$ \\
		$ \ \ \ \ 0.807_{-0.015}^{+0.019}$ \\
		$ \ \ \ \ 0.807_{-0.014}^{+0.019}$ \\
		$ \ \ \ \ 0.807_{-0.013}^{+0.019}$ \\
		$ \ \ \ \ 0.807_{-0.013}^{+0.019}$ \\
		$ \ \ \ \ \ \ \ \ - $
		\end{tabular}%
		$ & $%
		\begin{tabular}{l}
		$ \ \ -0.063_{-0.026}^{+0.038}$ \\
		$ \ \ -0.078_{-0.021}^{+0.034}$ \\
		$ \ \ -0.072 _{-0.022}^{+0.036}$ \\
		$ \ \ -0.069_{-0.024}^{+0.037}$ \\
		$ \ \ -0.068_{-0.024}^{+0.037}$ \\
		$ \ \ -0.067_{-0.025}^{+0.037}$ \\
		$ \ \ \ \ \ \ \ \ - $
		\end{tabular}%
		$ & $%
		\begin{tabular}{l}
		$ \ \ -0.311_{-0.008}^{+0.014}$ \\
		$ \ \ -0.314_{-0.002}^{+0.008}$ \\
		$ \ \ -0.314_{-0.004}^{+0.010}$ \\
		$ \ \ -0.314_{-0.005}^{+0.012}$ \\
		$ \ \ -0.314_{-0.006}^{+0.012}$ \\
		$ \ \ -0.313_{-0.006}^{+0.013}$ \\
		$ \ \ \ \ \ \ \ \ - $
		\end{tabular}%
		$ & $%
		\begin{tabular}{l}
		$ \ \ \ 0.021_{-0.009}^{+0.007}$ \\
		$ \ \ \ 0.024_{-0.009}^{+0.008}$ \\
		$ \ \ \ 0.023_{-0.009}^{+0.007}$ \\
		$ \ \ \ 0.023_{-0.009}^{+0.007}$ \\
		$ \ \ \ 0.022_{-0.009}^{+0.007}$ \\
		$ \ \ \ 0.022_{-0.009}^{+0.007}$ \\
		$ \ \ \ \ \ \ \ \ - $
		\end{tabular}
		$ \\ \hline
		$\lbrack 6,8]%
		\begin{tabular}{l}
		$\text{SM}$ \\
		$\text{RS}_c|_{M_{g^{(1)}}}=4.8$ \\
		$\text{RS}_c|_{M_{g^{(1)}}}=6.1$ \\
		$\text{RS}_c|_{M_{g^{(1)}}}=7.4$ \\
		$\text{RS}_c|_{M_{g^{(1)}}}=8.7$ \\
		$\text{RS}_c|_{M_{g^{(1)}}}=10$ \\
		$  \text{LHCb} $
		\end{tabular}%
		$ & $%
		\begin{tabular}{l}
		$ \ 0.312_{-0.094}^{+0.094}$ \\
		$ \ 0.306_{-0.093}^{+0.094}$ \\
		$ \ 0.307_{-0.093}^{+0.094}$ \\
		$ \ 0.308_{-0.093}^{+0.094}$ \\
		$ \ 0.309_{-0.094}^{+0.094}$ \\
		$ \ 0.310_{-0.094}^{+0.094}$ \\
		$ \ 0.25_{-0.111}^{+0.120}$
		\end{tabular}%
		$ & $%
		\begin{tabular}{l}
		$ \ \ \ \ 0.724_{-0.014}^{+0.025}$ \\
		$ \ \ \ \  0.720_{-0.016}^{+0.026}$ \\
		$ \ \ \ \ 0.721_{-0.016}^{+0.026}$ \\
		$ \ \ \ \ 0.722_{-0.015}^{+0.025}$ \\
		$ \ \ \ \ 0.723_{-0.015}^{+0.025}$ \\
		$ \ \ \ \ 0.723_{-0.014}^{+0.025}$ \\
		$ \ \ \ \ \ \ \ \ - $
		\end{tabular}%
		$ & $%
		\begin{tabular}{l}
		$ \ \ -0.162_{-0.017}^{+0.025}$ \\
		$ \ \ -0.174_{-0.013}^{+0.021}$ \\
		$ \ \ -0.170_{-0.014}^{+0.022}$ \\
		$ \ \ -0.168_{-0.015}^{+0.023}$ \\
		$ \ \ -0.166_{-0.016}^{+0.024}$ \\
		$ \ \ -0.165_{-0.016}^{+0.024}$ \\
		$ \ \ \ \ \ \ \ \ - $
		\end{tabular}%
		$ & $%
		\begin{tabular}{l}
		$ \ \ -0.317_{-0.004}^{+0.007}$ \\
		$ \ \  -0.314_{-0.001}^{+0.002}$ \\
		$ \ \  -0.317_{-0.001}^{+0.004}$ \\
		$ \ \  -0.317_{-0.002}^{+0.005}$ \\
		$ \ \  -0.317_{-0.002}^{+0.006}$ \\
		$ \ \  -0.317_{-0.003}^{+0.006}$ \\
		$ \ \ \ \ \ \ \ \ - $
		\end{tabular}%
		$ & $%
		\begin{tabular}{l}
		$ \ \ \ 0.052_{-0.007}^{+0.005}$ \\
		$ \ \ \ 0.054_{-0.007}^{+0.005}$ \\
		$ \ \ \ 0.054_{-0.007}^{+0.005}$ \\
		$ \ \ \ 0.054_{-0.007}^{+0.005}$ \\
		$ \ \ \ 0.053_{-0.007}^{+0.005}$ \\
		$ \ \ \ 0.053_{-0.007}^{+0.006}$ \\
		$ \ \ \ \ \ \ \ \ - $
		\end{tabular}
		$ \\ \hline
		$\lbrack 1.1,6]%
				\begin{tabular}{l}
					$\text{SM}$ \\
					$\text{RS}_c|_{M_{g^{(1)}}}=4.8$ \\
					$\text{RS}_c|_{M_{g^{(1)}}}=6.1$ \\
					$\text{RS}_c|_{M_{g^{(1)}}}=7.4$ \\
					$\text{RS}_c|_{M_{g^{(1)}}}=8.7$ \\
					$\text{RS}_c|_{M_{g^{(1)}}}=10$ \\
					$  \text{LHCb} $
				\end{tabular}%
				$ & $%
				\begin{tabular}{l}
				$ \ 0.199_{-0.120}^{+0.120}$ \\
				$ \ 0.190_{-0.119}^{+0.120}$ \\
				$ \ 0.193_{-0.119}^{+0.120}$ \\
				$ \ 0.195_{-0.119}^{+0.120}$ \\
				$ \ 0.196_{-0.119}^{+0.120}$ \\
				$ \ 0.197_{-0.120}^{+0.120}$ \\
				$ \ 0.09_{-0.051}^{+0.061}$
				\end{tabular}%
				$ & $%
				\begin{tabular}{l}
				$ \ \ \ \ \ 0.818_{-0.011}^{+0.011}$ \\
				$ \ \ \ \ \ 0.824_{-0.007}^{+0.010}$ \\
				$  \ \ \ \ \ 0.821_{-0.008}^{+0.010}$  \\
				$ \ \ \ \ \  0.820_{-0.010}^{+0.010}$ \\
				$ \ \ \ \ \ 0.819_{-0.010}^{+0.010}$ \\
				$ \ \ \ \ \ 0.819_{-0.011}^{+0.011}$ \\
				$ \ \ \ \ \ \ \ \ - $
				\end{tabular}%
				$ & $%
				\begin{tabular}{l}
				$ \ \ \ \ \ 0.009_{-0.018}^{+0.027}$ \\
				$ \ \ \  -0.005_{-0.014}^{+0.025}$ \\
				$ \ \ \ \ \ 0.001_{-0.015}^{+0.026}$ \\
				$ \ \ \ \ \ 0.003_{-0.016}^{+0.026}$ \\
				$ \ \ \ \ \ 0.005_{-0.016}^{+0.026}$ \\
				$ \ \ \ \ \ 0.006_{-0.017}^{+0.026}$ \\
				$ \ \ \ \ \ \ \ \ - $
				\end{tabular}%
				$ & $%
				\begin{tabular}{l}
				$ \ \  -0.309_{-0.010}^{+0.018}$ \\
				$ \ \   -0.312_{-0.004}^{+0.012}$ \\
				$  \ \  -0.312_{-0.006}^{+0.014}$ \\
				$  \ \  -0.312_{-0.007}^{+0.016}$ \\
				$ \ \   -0.311_{-0.008}^{+0.016}$ \\
				$ \ \   -0.311_{-0.008}^{+0.017}$ \\
				$ \ \ \ \ \ \ \ \ \ - $
				\end{tabular}%
				$ & $%
				\begin{tabular}{l}
				$ \ \ -0.002_{-0.005}^{+0.004}$ \\
				$ \ \ \ \ 0.001_{-0.006}^{+0.005}$ \\
				$ \ \ \ \ 0.000_{-0.006}^{+0.005}$ \\
				$ \ \ -0.001_{-0.005}^{+0.005}$ \\
				$ \ \ -0.001_{-0.005}^{+0.005}$ \\
				$ \ \ -0.001_{-0.005}^{+0.004}$ \\
				$ \ \ \ \ \ \ \  - $
				\end{tabular}
				$ \\ \hline
	\end{tabular}%
\end{scriptsize}
\end{table*}

Next, we compare our results of observables in the SM and the $\text{RS}_c$ model with the measurements from the LHCb experiment \cite{Aaij:2015xza}. For most of the observables, results in the $\text {RS}_c$ model are close to that obtained for the SM in all bins of $s$ and this can be seen in Tables \ref{low-s-bin-values} and \ref{high-s-bin-values}. The branching ratio for the four body decay process $\Lambda_b \rightarrow \Lambda ( \rightarrow p \pi ) \mu^{+} \mu^-$ in the $\text{RS}_c$ model (for $M_{g^{(1)}}=4.8$ TeV) shows a slight deviation at low recoil and almost no deviation at large recoil. For the bin $[1.1,6]$, the branching ratio in the SM and the $\text {RS}_c$ are $ 0.199_{-0.12}^{+0.12}$  and $ 0.190_{-0.119}^{+0.120}$ respectively which are $ 1.8 \sigma$ and $ 1.9 \sigma$ away from the measured value $  0.09_{-0.051}^{+0.061}$. The situation is quite similar for all other bins of large recoil where values of observables do not change much even for $M_{g^{(1)}}=4.8$ TeV. For low recoil bin $[15,20]$, the SM and the $\text {RS}_c$ model results
$0.753_{-0.069}^{+0.069}$ and $ 0.807_{-0.069}^{+0.069} $ deviate from the measured value by $ 4.7 \sigma $ and $ 4.1 \sigma$. It is noted that the
differential branching ratio in the $\text{RS}_c$ model is lower than the SM at large recoil and higher than the SM at low recoil. \\

\begin{table*}
\caption{Numerical results of the observables (high $s$ region) in the $\Lambda_b \rightarrow \Lambda (\rightarrow p \pi) \mu^+ \mu^-$ decay,
 obtained for the SM and the $\text{RS}_c$ model with $y_{\star} =1.5$ case, in different bins of high $s$. Experimentally measured values are taken from \cite{Aaij:2015xza}.} \label{high-s-bin-values}
\begin{scriptsize}
\begin{tabular}{l||llllll}
\hline\hline
& $\ \left\langle \frac{d\beta }{ds}\times 10^{-7}\right\rangle $ & $\ \ \ \
\ \ \left\langle F_{L}\right\rangle $  & $\ \ \ \ \ \ \left\langle
A_{FB}^{\ell }\right\rangle $ & $\ \ \ \  \left\langle
A_{FB}^{\Lambda}\right\rangle $ & $\ \ \  \left\langle A_{FB}^{l\Lambda
}\right\rangle
$ \\ \hline
\hline	
		$\lbrack 15,16]%
		\begin{tabular}{l}
		$\text{SM}$ \\
		$\text{RS}_c|_{M_{g^{(1)}}}=4.8$ \\
		$\text{RS}_c|_{M_{g^{(1)}}}=6.1$ \\
		$\text{RS}_c|_{M_{g^{(1)}}}=7.4$ \\
		$\text{RS}_c|_{M_{g^{(1)}}}=8.7$ \\
		$\text{RS}_c|_{M_{g^{(1)}}}=10$ \\
		$\text{LHCb}$ \\
		\end{tabular}%
		$ & $%
		\begin{tabular}{l}
		$ \ 0.798_{-0.073}^{+0.073}$ \\
		$ \ 0.832_{-0.073}^{+0.073}$ \\
		$ \ 0.816_{-0.073}^{+0.073}$  \\
		$ \ 0.810_{-0.073}^{+0.073}$  \\
		$ \ 0.806_{-0.074}^{+0.074}$  \\
		$ \ 0.804_{-0.074}^{+0.074}$  \\
		$ \ 1.12_{-0.187}^{+0.197}$
		\end{tabular}%
		$ & $%
		\begin{tabular}{l}
		$ \ \ \ \ \ 0.454_{-0.017}^{+0.032}$ \\
		$ \ \ \ \ \ 0.447_{-0.017}^{+0.033}$ \\
		$ \ \ \ \ \  0.450_{-0.017}^{+0.033}$ \\
		$ \ \ \ \ \ 0.451_{-0.017}^{+0.032}$ \\
		$ \ \ \ \ \ 0.452_{-0.017}^{+0.032}$ \\
		$ \ \ \ \ \ 0.452_{-0.017}^{+0.032}$ \\
		$ \ \ \ \ \ 0.49_{-0.304}^{+0.304}$
		\end{tabular}%
		$ & $%
		\begin{tabular}{l}
		$ \ \ \ -0.382_{-0.008}^{+0.017}$ \\
		$ \ \ \ -0.365_{-0.006}^{+0.014}$ \\
		$ \ \ \ -0.372_{-0.007}^{+0.015}$ \\
		$ \ \ \ -0.375_{-0.007}^{+0.015}$ \\
		$ \ \ \ -0.377_{-0.007}^{+0.016}$ \\
		$ \ \ \ -0.378_{-0.008}^{+0.016}$  \\
		$ \ \ \  -0.10_{-0.163}^{+0.183}$
		\end{tabular}%
		$ & $%
		\begin{tabular}{l}
		$ \ \ -0.307_{-0.004}^{+0.002}$ \\
		$ \ \   -0.287_{-0.005}^{+0.003}$ \\
		$ \ \   -0.296_{-0.005}^{+0.003}$ \\
		$  \ \  -0.300_{-0.004}^{+0.003}$ \\
		$  \ \  -0.302_{-0.004}^{+0.003}$ \\
		$  \ \  -0.304_{-0.004}^{+0.002}$ \\
		$ \ \  -0.19_{-0.163}^{+0.143}$
		\end{tabular}%
		$ & $%
		\begin{tabular}{l}
		$ \ \ \ \ 0.131_{-0.008}^{+0.004}$ \\
		$ \ \ \ \ 0.132_{-0.008}^{+0.004}$ \\
		$ \ \ \ \ 0.132_{-0.008}^{+0.004}$ \\
		$ \ \ \ \ 0.132_{-0.008}^{+0.004}$ \\
		$ \ \ \ \ 0.132_{-0.008}^{+0.004}$ \\
		$ \ \ \ \ 0.132_{-0.008}^{+0.004}$ \\
		$ \ \ \ \ \ \ \ \ - $
		\end{tabular}
		$ \\ \hline
		$\lbrack 16,18]%
		\begin{tabular}{l}
		$\text{SM}$ \\
		$\text{RS}_c|_{M_{g^{(1)}}}=4.8$ \\
		$\text{RS}_c|_{M_{g^{(1)}}}=6.1$ \\
		$\text{RS}_c|_{M_{g^{(1)}}}=7.4$ \\
		$\text{RS}_c|_{M_{g^{(1)}}}=8.7$ \\
		$\text{RS}_c|_{M_{g^{(1)}}}=10$ \\
		$ \text{LHCb} $
		\end{tabular}%
		$ & $%
		\begin{tabular}{l}
		$ \ 0.825_{-0.075}^{+0.075}$ \\
		$ \ 0.877_{-0.075}^{+0.075}$ \\
		$ \ 0.855_{-0.075}^{+0.075}$ \\
		$ \ 0.844_{-0.075}^{+0.075}$ \\
		$ \ 0.838_{-0.075}^{+0.075}$ \\
		$ \ 0.835_{-0.075}^{+0.075}$ \\
		$ \ 1.22_{-0.152}^{+0.143}$
		\end{tabular}%
		$ & $%
		\begin{tabular}{l}
		$ \ \ \ \ \  0.418_{-0.017}^{+0.033}$ \\
		$ \ \ \ \ \ 0.411_{-0.017}^{+0.033}$ \\
		$ \ \ \ \ \ 0.414_{-0.017}^{+0.033} $ \\
		$ \ \ \ \ \ 0.415_{-0.017}^{+0.033} $ \\
		$ \ \ \ \ \ 0.416_{-0.017}^{+0.033} $  \\
		$ \ \ \ \ \ 0.416_{-0.017}^{+0.033} $ \\
		$ \ \ \ \ \ 0.68_{-0.216}^{+0.158}$
		\end{tabular}%
		$ & $%
		\begin{tabular}{l}
		$  \ \ \  -0.381_{-0.006}^{+0.013}$ \\
		$ \ \ \ -0.356_{-0.004}^{+0.010}$ \\
		$ \ \ \  -0.366_{-0.005}^{+0.011}$ \\
		$ \ \ \  -0.371_{-0.005}^{+0.012}$ \\
		$ \ \ \ -0.374_{-0.005}^{+0.012}$ \\
		$ \ \ \ -0.376_{-0.006}^{+0.012}$ \\
		$ \ \ \ -0.07_{-0.127}^{+0.136}$
		\end{tabular}
		$ & $%
		\begin{tabular}{l}
		$  \ \  -0.289_{-0.006}^{+0.005}$ \\
		$  \ \  -0.265_{-0.006}^{+0.005}$ \\
		$  \ \  -0.276_{-0.006}^{+0.005}$ \\
		$  \ \  -0.280_{-0.006}^{+0.005}$ \\
		$  \ \  -0.283_{-0.006}^{+0.005}$ \\
		$  \ \  -0.284_{-0.006}^{+0.005}$ \\
		$  \ \  -0.44_{-0.058}^{+0.104}$
		\end{tabular}%
		$ & $%
		\begin{tabular}{l}
		$ \ \ \ \ 0.141_{-0.008}^{+0.004}$ \\
		$ \ \ \ \ 0.140_{-0.009}^{+0.004}$ \\
		$ \ \ \ \  0.141_{-0.008}^{+0.004}$ \\
		$ \ \ \ \ 0.141_{-0.008}^{+0.004}$ \\
		$ \ \ \ \ 0.141_{-0.008}^{+0.004}$ \\
		$ \ \ \ \  0.141_{-0.008}^{+0.004}$ \\
		$ \ \ \ \ \ \ \ \ -  $
		\end{tabular}
		$ \\ \hline
		$\lbrack 18,20]%
		\begin{tabular}{l}
		$\text{SM}$ \\
		$\text{RS}_c|_{M_{g^{(1)}}}=4.8$ \\
		$\text{RS}_c|_{M_{g^{(1)}}}=6.1$ \\
		$\text{RS}_c|_{M_{g^{(1)}}}=7.4$ \\
		$\text{RS}_c|_{M_{g^{(1)}}}=8.7$ \\
		$\text{RS}_c|_{M_{g^{(1)}}}=10$ \\
		$ \text{LHCb} $
		\end{tabular}%
		$ & $%
		\begin{tabular}{l}
		$  \ 0.658_{-0.066}^{+0.066}$ \\
		$  \ 0.726_{-0.066}^{+0.066}$ \\
		$ \ 0.698_{-0.066}^{+0.066}$ \\
		$ \ 0.685_{-0.066}^{+0.066}$ \\
		$ \ 0.677_{-0.066}^{+0.066}$ \\
		$ \ 0.672_{-0.066}^{+0.066}$ \\
		$ \ 1.24_{-0.149}^{+0.152}$
		\end{tabular}%
		$ & $%
		\begin{tabular}{l}
		$ \ \ \ \ \ 0.371_{-0.019}^{+0.034}$ \\
		$ \ \ \ \ \ 0.367_{-0.020}^{+0.034}$ \\
		$ \ \ \ \ \ 0.368_{-0.019}^{+0.034}$ \\
		$ \ \ \ \ \ 0.369_{-0.019}^{+0.034}$ \\
		$ \ \ \ \ \ 0.370_{-0.019}^{+0.034}$ \\
		$ \ \ \ \ \ 0.370_{-0.019}^{+0.034}$ \\
		$ \ \ \ \ \ 0.62_{-0.273}^{+0.243}$
		\end{tabular}%
		$ & $%
		\begin{tabular}{l}
		$  \ \ \ -0.317_{-0.010}^{+0.010}$  \\
		$ \ \ \ -0.286_{-0.010}^{+0.010}$ \\
		$ \ \ \  -0.297_{-0.010}^{+0.010}$ \\
		$ \ \ \ -0.303_{-0.010}^{+0.010}$ \\
		$ \ \ \ -0.307_{-0.010}^{+0.010}$ \\
		$ \ \ \ -0.309_{-0.010}^{+0.010}$ \\
		$ \ \ \ \ \ 0.01_{-0.146}^{+0.155}$
		\end{tabular}%
		$ & $%
		\begin{tabular}{l}
		$ \ \ -0.227_{-0.011}^{+0.011}$ \\
		$ \ \ -0.201_{-0.010}^{+0.010}$ \\
		$ \ \ -0.211_{-0.010}^{+0.010}$ \\
		$ \ \ -0.216_{-0.011}^{+0.011}$ \\
		$ \ \ -0.219_{-0.011}^{+0.011}$ \\
		$ \ \ -0.221_{-0.011}^{+0.011}$  \\
		$ \ \ -0.13_{-0.124}^{+0.095}$
		\end{tabular}%
		$ & $%
		\begin{tabular}{l}
		$ \ \ \ 0.153_{-0.009}^{+0.005}$ \\
		$ \ \ \ 0.151_{-0.009}^{+0.005}$ \\
		$ \ \ \ 0.152_{-0.009}^{+0.005}$ \\
		$ \ \ \ 0.152_{-0.009}^{+0.005}$ \\
		$ \ \ \ 0.153_{-0.009}^{+0.005}$ \\
		$ \ \ \ 0.153_{-0.009}^{+0.005}$  \\
		$ \ \ \ \ \ \ \ - \ \ $
		\end{tabular}
		$ \\ \hline
		$\lbrack 15,20]%
		\begin{tabular}{l}
		$\text{SM}$ \\
		$\text{RS}_c|_{M_{g^{(1)}}}=4.8$ \\
		$\text{RS}_c|_{M_{g^{(1)}}}=6.1$ \\
		$\text{RS}_c|_{M_{g^{(1)}}}=7.4$ \\
		$\text{RS}_c|_{M_{g^{(1)}}}=8.7$ \\
		$\text{RS}_c|_{M_{g^{(1)}}}=10$ \\
		$ \text{LHCb} $
		\end{tabular}%
		$ & $%
		\begin{tabular}{l}
		$ \ 0.753_{-0.069}^{+0.069}$ \\
		$ \ 0.807_{-0.069}^{+0.069}$ \\
		$ \ 0.785_{-0.069}^{+0.069}$ \\
		$ \ 0.774_{-0.069}^{+0.069}$  \\
		$ \ 0.767_{-0.069}^{+0.069}$  \\
		$ \ 0.764_{-0.069}^{+0.069}$  \\
		$ \ 1.20_{-0.099}^{+0.092}$
		\end{tabular}%
		$ & $%
		\begin{tabular}{l}
		$ \ \ \ \ \  0.409_{-0.018}^{+0.033}$ \\
		$ \ \ \ \ \  0.403_{-0.019}^{+0.034}$ \\
		$ \ \ \ \ \ 0.405_{-0.019}^{+0.034}$ \\
		$ \ \ \ \ \ 0.406_{-0.019}^{+0.033}$ \\
		$ \ \ \ \ \ 0.407_{-0.019}^{+0.033}$ \\
		$ \ \ \ \ \ 0.407_{-0.019}^{+0.033}$ \\
		$ \ \ \ \ \ 0.61_{-0.143}^{+0.114}$
		\end{tabular}%
		$ & $%
		\begin{tabular}{l}
		$ \ \ \ -0.358_{-0.007}^{+0.012}$ \\
		$ \ \ \ -0.332_{-0.009}^{+0.008}$ \\
		$ \ \ \ -0.343_{-0.008}^{+0.010}$ \\
		$ \ \ \ -0.348_{-0.007}^{+0.010}$ \\
		$ \ \ \ -0.351_{-0.007}^{+0.011}$ \\
		$ \ \ \ -0.353_{-0.007}^{+0.011}$ \\
		$ \ \ \ -0.05_{-0.095}^{+0.095}$
		\end{tabular}%
		$ & $%
		\begin{tabular}{l}
		$ \ \ -0.271_{-0.011}^{+0.011}$ \\
		$ \ \ -0.247_{-0.011}^{+0.011}$ \\
		$ \ \ -0.257_{-0.011}^{+0.011}$ \\
		$ \ \ -0.262_{-0.011}^{+0.011}$ \\
		$ \ \ -0.264_{-0.011}^{+0.011}$ \\
		$ \ \ -0.266_{-0.011}^{+0.011}$ \\
		$ \ \ -0.29_{-0.081}^{+0.076}$
		\end{tabular}%
		$ & $%
		\begin{tabular}{l}
		$ \ \ \ 0.143_{-0.008}^{+0.005}$ \\
		$ \ \ \ 0.142_{-0.009}^{+0.005}$ \\
		$ \ \ \ 0.143_{-0.009}^{+0.005}$ \\
		$ \ \ \ 0.143_{-0.009}^{+0.005}$ \\
		$ \ \ \ 0.143_{-0.009}^{+0.005}$ \\
		$ \ \ \ 0.143_{-0.008}^{+0.005}$ \\
		$ \ \ \ \ \ \ \ - \ \ \ $
		\end{tabular}
		$ \\ \hline
	\end{tabular}%
\end{scriptsize}
\end{table*}
In case of $F_L$, maximum deviation has been observed for the first bin $[0.1,2]$ GeV$^2$ where predictions in the SM and the $\text {RS}_c$ model are $ \left\langle F_{L}\right\rangle_{SM} =0.535^{+0.065}_{-0.078}$ and $ \left\langle F_{L}\right\rangle_{\text {RS}_c} =0.552_{-0.084}^{+0.069}$, respectively which vary from the measured value $0.56_{-0.566}^{+0.244}$ by $ 0.1 \sigma$ and  $ 0.02 \sigma$, respectively. For most of the bins, deviation of $F_L$ in the $\text {RS}_c$ model from the SM is negligible. For low recoil bin $[15,20]$ GeV$^2$, the values in both models $ \left\langle F_{L}\right\rangle_{SM}=0.409^{+0.033}_{-0.018}$, $ \left\langle F_{L}\right\rangle_{\text {RS}_c} =0.403_{-0.019}^{+0.034}$ deviate from the experimental result $0.61^{+0.114}_{-0.143}$ in the same bin by $ 1.6 \sigma $. At
lower values of $s$ upto $4$ GeV$^2$, the $ \text {RS}_c $ model results deviate from the SM values to a greater extent, whereas almost similar values of the
$ \text {RS}_c $ model are obtained for the rest of the spectrum.

For $A^{\ell}_{FB}$, small deviation in the $\text {RS}_c$ model exists from the SM  at low recoil. In the first bin $[0.1,2]$ GeV$^2$our calculated results in both models differ from the measured value by $0.6 \sigma$. For large $s$ bin $[15,20]$ GeV$^2$, the values in both models $\left\langle A^{\ell}_{FB}\right\rangle_{\text SM} =-0.358_{-0.007}^{+0.012}$ and $ \left\langle A^{\ell}_{FB} \right\rangle_{\text {RS}_c} =-0.332_{-0.009}^{+0.008} $ are very close to each other and are $ 3.2 \sigma$ and $ 3.0 \sigma $ away from the measured value $-0.05_{-0.095}^{+0.095}$ in the same bin.

For $A^{\Lambda}_{FB}$ in the bin $[15,20]$ GeV$^2$ results of the SM and the $ \text {RS}_c $ model are $ \left\langle A^{\Lambda}_{FB} \right\rangle_{\text SM}= -0.271_{-0.011}^{+0.011}$ and $ \left\langle A^{\Lambda}_{FB} \right\rangle_{\text {RS}_c}= -0.247_{-0.011}^{+0.011} $ and deviate from the measured value of LHCb  $ -0.29_{-0.081}^{+0.076}$ by $ 0.2 \sigma $ and $ 0.5 \sigma $.
For  $A^{\ell \Lambda}_{FB}$, no sizable deviation from the SM has been observed in any $s$ bin for the $\text {RS}_c$ model.

\section{Conclusions\label{conc}}
In the work presented here, we have studied the angular observables of the theoretically clean decay $\Lambda_b \rightarrow \Lambda (\to p \pi^{-}) \mu^{+} \mu^{-} $ in the SM
and the Randall-Sundrum model with custodial protection. After performing the scan of the parameter space of the model in the light of current constraints,
we have worked out the largest possible deviations in the Wilson coefficients $|\Delta C_{7,9,10}^{(\prime)}|$ from the SM predictions for different allowed
values of KK gluon mass $M_{g^{(1)}}$. The resultant deviations are small and do not allow for large effects in the angular observables.
Although for maximum possible deviations in Wilson coefficients, for $M_{g^{(1)}}=4.8$ TeV, in the $\text {RS}_c$ model, some of the observables receive considerable change in particular bins such as $\frac{d\mathcal{B}}{ds}$ and $A^{\ell}_{FB}$ in low recoil bin $[15,20]$ GeV$^2$ and $F_L$ in the bin $[0.1,2]$ GeV$^2$ but these deviations are still small to explain the large gap between the theoretical and experimental data. Therefore, it is concluded that under the present bounds on the mass of first KK gluon state $M_{g^{(1)}}$, observables are largely unaffected by the NP arising due to custodially protected RS model. Hence, the current constraints on the parameters of $\text{RS}_c$ are too strict to explain the observed deviations in different observables of $\Lambda_b \rightarrow \Lambda (\to p \pi^{-}) \mu^{+} \mu^{-}$ decay.

\section*{Acknowledgements}
F.M.B would like to acknowledge financial support from CAS-TWAS president's fellowship program. The work of F.M.B is also partly supported by National Science Foundation of China (11521505, 11621131001) and that of M.J.A by the URF (2015). In addition, A.N. would like to thank Dr. Zaheer Asghar for his help in the calculations done during this work.

\bibliographystyle{refstyle}
\bibliography{bibfile}
\end{document}